\documentclass[aps,pra,reprint,floatfix,showpacs,superscriptaddress]{revtex4-1}
\usepackage{hyperref}
\usepackage{amssymb}
\usepackage{amsmath}
\usepackage{amsfonts}
\usepackage{amsthm}
\usepackage{mathtools}
\usepackage{tensor}
\usepackage{mathrsfs}
\usepackage{latexsym}
\usepackage{theoremref}
\usepackage{paralist}
\usepackage{graphicx}
\usepackage{braket}
\usepackage{dsfont}
\usepackage{dcolumn}
\usepackage{array}
\usepackage{natbib}
\usepackage{cleveref}
\usepackage{bm}
\usepackage{color}

\newcommand{\dd}{\, \mathrm{d}}
\renewcommand{\vec}[1]{\mathbf #1}

\newcommand{\e}{\mathrm{e}}
\renewcommand{\i}{\mathrm{i}}
\crefname{equation}{Eq.}{Eqs.}
\crefname{figure}{Fig.}{Figs.}
\Crefname{figure}{Figure}{Figures}

\allowdisplaybreaks

\begin{document}
\title{Diagrammatic Approach to Multiphoton Scattering}
\author{Tian Feng See}
\email{baise.feng@gmail.com}
\affiliation{Centre for Quantum Technologies, National University of Singapore, 3 Science Drive 2, Singapore 117543}
\author{Changsuk Noh}
\email{undefying@gmail.com}
\affiliation{Korea Institute for Advanced Study, 85 Hoegiro, Seoul 130-722, Korea}
\author{Dimitris G. Angelakis}
\email{dimitris.angelakis@gmail.com}
\affiliation{Centre for Quantum Technologies, National University of Singapore, 3 Science Drive 2, Singapore 117543}
\affiliation{School of Electronic and Computer Engineering, Technical University of Crete, Chania, Greece 73100}
\date{\today}

\begin{abstract}
We present a  method to systematically study multi-photon transmission in one dimensional systems comprised of correlated quantum emitters coupled to input and output waveguides. Within the Green's function approach of the scattering matrix (S-matrix), we develop a diagrammatic technique to analytically obtain the system's scattering amplitudes while at the same time visualise all the possible absorption and emission processes. Our method helps to reduce the significant effort in finding the general response of a many-body bosonic system, particularly the nonlinear response embedded in the Green's functions. We demonstrate our proposal through physically relevant examples involving scattering of multi-photon states from two-level emitters as well as from arrays of correlated Kerr nonlinear resonators in the Bose-Hubbard model.
\end{abstract}

\maketitle

\section{Introduction}
Engineering strong photon-photon interactions \citep{chang14} is important in numerous areas such as quantum information processing and computing \citep{ultracoldatoms}. To achieve that, strong coupling between light and matter at the quantum level is often required. For this matter, different approaches have been explored using a variety of quantum technological platforms including atoms in optical and microwave cavities \citep{exploring}, semiconductor based devices \citep{semiconductor}, superconducting circuits \citep{wallraff04}, and Rydberg media \citep{firstenberg16}.  

Among other applications, strong photon-photon interactions allow generation and manipulation of non-classical states of light with applications in single photon transistors and quantum photonic switches \citep{chang07,michler00,volz12,tiecke14} . It can also be used to create strongly correlated states of light with applications in quantum simulations \cite{noh16,hartmann16}.

To determine if an atomic or a general quantum optical system can be used to manipulate light as above, one usually analyses its transmission spectra and photon statistics. Different theoretical methods have been used to retrieve such information in various setups. Since we are dealing with photons, it is natural to use quantum optical methods like master equations and the input-output formalism \citep{quantumnoise,statsmeth} to connect experimental observables such as the intensity and correlation functions of the transmitted or scatteredred light to the behaviour of the system probed. Tools that are not traditionally from quantum optics like Lippmann-Schwinger equation \citep{lippmann50} or equivalently Bethe ansatz \citep{bethe31} and quantum field theory \citep{introtoqft}, have also been incorporated recently when one is seeking exact analytical descriptions \citep{shen07,shi09,pletyukhov12,*schneider16,*kocabas16,zheng12,lee15,shi15,caneva15,xu15,roulet16}. For the case of photonic Fock state inputs, while it is relatively easy to calculate the part of the S-matrix due to elastic scattering, where the momenta of the photons are simply rearranged at the output, it is usually tedious and complicated to calculate the part due to inelastic scattering, where the output momenta are a continuum. The latter is particularly true for cases where more than two photons and/or several emitters are involved. 

One of the main focuses in this paper is to provide a systematic method to calculate these non-trivial terms. To do so, we use input-output operators to define the S-matrix and with inspiration from quantum field theory, we develop a diagrammatic method to evaluate the S-matrix in a setup where single-mode waveguides are coupled locally to an arbitrary system. Using this method, scattering elements are calculated more intuitively, especially their inelastic parts representing the nonlinear responses. We first present the model and the necessary background of S-matrix in \cref{sec:model}. Next, in \cref{sec:diagram}, we illustrate rules for drawing diagrams which allow their associated Green's functions to be written. We demonstrate our method with a few concrete examples in \cref{sec:examples} before concluding the paper. In particular, we will first discuss two non-interacting systems --- a two-level emitter and two collocated two-level emitters, before moving on to a many emitters interacting system described by the Bose-Hubbard model. In the latter, we consider  open and closed boundary conditions. Through these examples, we aim to illuminate how the diagrammatic approach can help one understand the physics of the systems both qualitatively and quantitatively. 
 
\section{Model and background of S-matrix}\label{sec:model}
\begin{figure}[!h]
\centering
\includegraphics[width=0.4\textwidth]{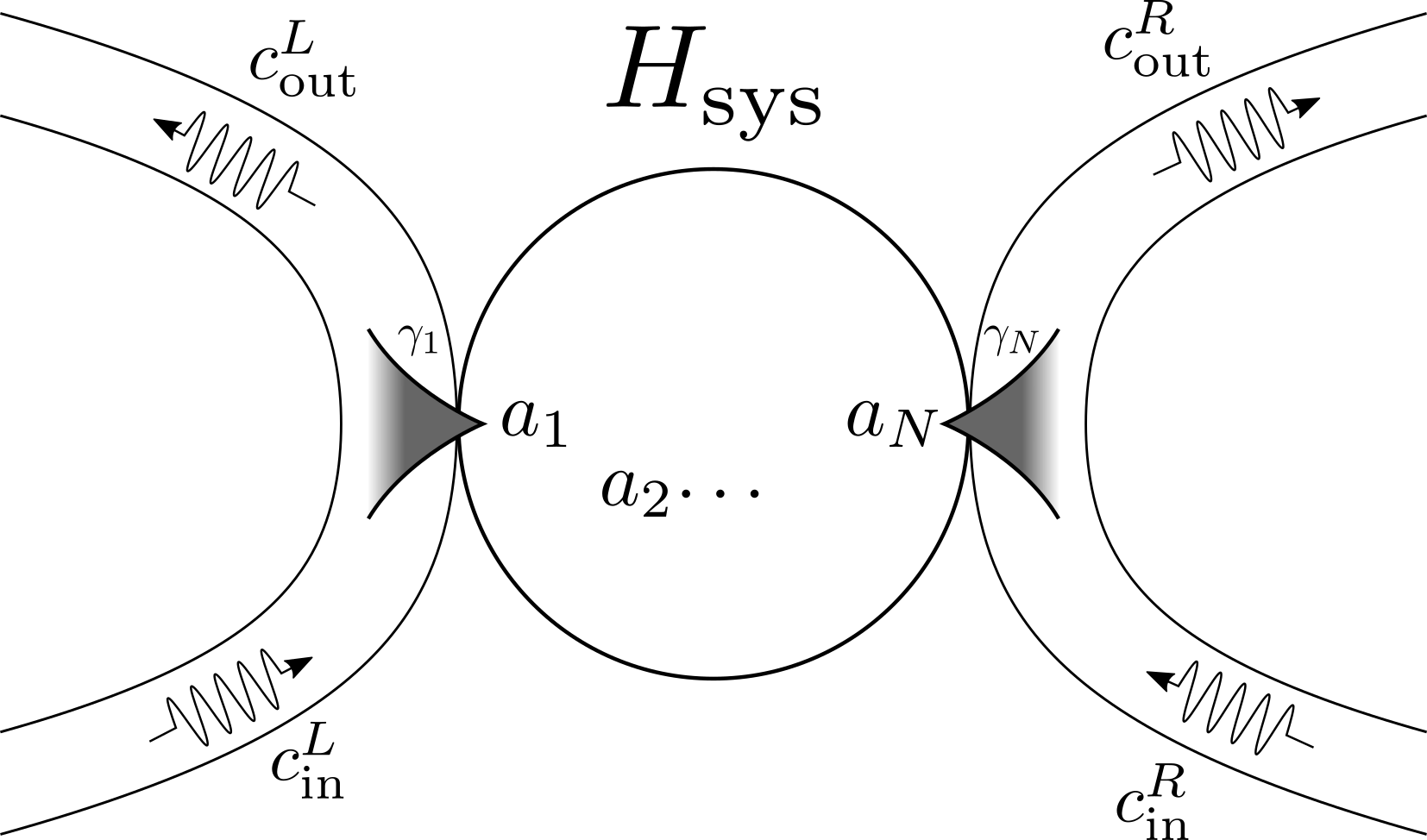}\\
\caption{Two photonic waveguides coupled to a many-body system with Hamiltonian, $H_\text{sys}$ that is described by the operators $a_1, \dots, a_N$. The left and right waveguides are bilinearly coupled to $a_1$ and $a_N$ respectively.}\label{fig:setup}
\end{figure}

To study the transport and scattering properties of a system, we couple single-mode waveguides to the system (\cref{fig:setup}), insert photons through the waveguides, and observe the reflection and transmission spectra. Such a setup is described by a total Hamiltonian ($\hbar = 1$)
\[H_\text{tot} = H_\text{w} + H_\text{ws} + H_\text{sys}.\]
The free propagation of photons in the left and right waveguides is described by
\[H_\text{w} = \int\dd k\; \omega(k)\left(c_k^{L\dagger} c_k^{L\vphantom{\dagger}} + c_k^{R\dagger} c_k^{R\vphantom{\dagger}}\right)\]
where $c^{L/R\dagger}_k$ and $c^{L/R}_k$ are the creation and annihilation operators of $k$-mode photons in the left/right waveguides respectively. We assume linear dispersion relation in the waveguides and take the group velocity to be one, i.e. $\omega(k) = k$. This allows us to use momentum and energy interchangeably throughout this paper. 

Next, we model the coupling between the waveguides and local system operators $a_1$ and $a_N$ as
\begin{equation}
H_\text{ws} = \int\dd k \left(\xi^{\vphantom{\dagger}}_1 a_1^\dagger c^L_k + \xi^{\vphantom{\dagger}}_N a_N^\dagger c^R_k + h.c.\right).
\end{equation}
We have made the rotating wave approximation and assumed frequency-independent coupling constants $\xi_1$ and $\xi_N$, which is valid when the coupling constant is small compared to the typical frequency of the system and the bandwidths of the waveguides are large. This assumption, along with the linear dispersion relation of the waveguides, is equivalent to making the Markov approximation, yielding the input-output relation to be derived shortly.

Finally, $H_\text{sys}$ is the Hamiltonian of the system of interest and has a form $H_\text{sys} = H_0 + H_\text{int}$, where $H_0 = \sum_{j = 1}^N \omega_j a_j^\dagger a_j$ describes the free energy of all its constituents, $a_j$'s and $H_\text{int}$ describes interactions between them. We restrict our discussion to interactions where the system Hamiltonian is total-particle-number conserving, i.e. $[H_\text{sys}, N_\text{tot}] = 0$, where $N_\text{tot} = \text{total particle number} = \sum_{j = 1}^N a_j^\dagger a_j$. Such feature is common in many physical models. Generalisation of the method to non-particle-number-conserving interactions is straightforward. 

From the above model Hamiltonian, one can derive the following input-output relations for the waveguides \citep{gardiner85},
\begin{align}\label{eq:inout}
c^L_\text{out}(t) &= c^L_\text{in}(t) - \i\sqrt{\gamma_1}a_1(t) \nonumber\\
c^R_\text{out}(t) &= c^R_\text{in}(t) - \i\sqrt{\gamma_N}a_N(t)
\end{align}
where $\gamma_i = 2\pi\xi_i^2$, for $i = 1, N$ and $c^{L/R}_\text{in}$ and $c^{L/R}_\text{out}$ are the input and output operators in the left/right waveguides. Using these operators, the S-matrix, which is a unitary matrix mapping the asymptotic free initial state $\ket{\vec{k}} = \prod_i c_{k_i}^\dagger\ket{0}$ to the asymptotic free final state $\ket{\vec{p}} = \prod_i c_{p_i}^\dagger\ket{0}$ with incoming and outgoing particle momenta of $\vec{k}$ and $\vec{p}$, is as follows,  
\begin{align*}
S(\vec{p}; \vec{k}) &= \braket{\vec{p}|S|\vec{k}} \\*
&= \mathscr{F} \braket{0|c_\text{out}(t'_1)\dots c_\text{out}(t'_n)c_\text{in}^\dagger(t_1)\dots c_\text{in}^\dagger(t_n)|0}
\end{align*}
where $\mathscr{F} = \left(\prod_{i = 1}^n\int\frac{\dd t'_i}{\sqrt{2\pi}}\e^{\i p_it'_i}\prod_{j = 1}^n\int\frac{\dd t_j}{\sqrt{2\pi}}\e^{-\i k_jt_j}\right)$ is the Fourier transform operation. We have omitted the labels for the incoming and outgoing photon paths, i.e. the $L$ and $R$ labels, on the input and output operators \footnote{For example, for a one-photon process, there are four different S-matrix elements for incoming and outgoing photons from the left and right waveguides, i.e.~$\tensor[_L]{\braket{p|S|k}}{_L}$, $\tensor[_R]{\braket{p|S|k}}{_L}$, $\tensor[_L]{\braket{p|S|k}}{_R}$ and $\tensor[_R]{\braket{p|S|k}}{_R}$.}.

As noted in Ref. \citep{xu15}, the S-matrix can be cluster decomposed into a sum of the products of connected parts, i.e.
\begin{equation}\label{eq:clusterd}
S_{\vec{p};\vec{k}} = \sum_\mathcal{B}\sum_P\prod_{i = 1}^{M_\mathcal{B}}S^C_{p_{\mathcal{B}_i};k_{P\mathcal{B}_i}}
\end{equation}
where the summation is taken over all partitions $\mathcal{B}$ of $\{1, 2, \dots, n\}$, with $M_\mathcal{B}$ number of subsets $\mathcal{B}_i$, and all permutations (denoted $P$) of the subset elements $P\mathcal{B}_i$. Note that the connected part of a function is defined as the part of the function that is proportional to only one delta function. Furthermore, a connected $n$-photon S-matrix, $S^C_{\vec{p};\vec{k}}$ is equal to a connected $2n$-point Green's function for $n > 1$. For $n = 1$, there is an extra delta function if there is an input from that channel (\cref{sec:clusterS}). Omitting the incoming and outgoing photon paths, the $2n$-point Green's function is defined as
\begin{align}\label{eq:gfun}
&G(p_1\dots p_n; k_1\dots k_n)\nonumber\\*
= &(-1)^n\mathscr{F}\Bra{0}\mathcal{T}a(t'_1)\dots a(t'_n)a^\dagger(t_1)\dots a^\dagger(t_n)\Ket{0}
\end{align}
where $\mathcal{T}$ is the time ordering operator and all $\gamma$'s in the definition have been set to one. The correct factor of $\gamma$'s could be recovered easily as each $a_j$ contributes a factor of $\sqrt{\gamma_j}$. The system operators evolve under the effective Hamiltonian,
\[H_\text{eff} = H_\text{sys} - \i\frac{\gamma_1}{2}a_1^\dagger a^{}_1 -\i\frac{\gamma_N}{2}a_N^\dagger a^{}_N\]
as
\[a(t) = \e^{iH_\text{eff}t}a\e^{-iH_\text{eff}t}, \qquad a^\dagger(t) = \e^{iH_\text{eff}t}a^\dagger \e^{-iH_\text{eff}t}.\]
For $n > 1$, the $2n$-point Green's function defined above is nonzero only when the system is nonlinear. It represents inelastic scattering processes when multiple photons are sent into the system. Elastic scattering processes, on the other hand, are fully described by products of single-photon S-matrix elements. 

In summary, to compute the S-matrix, we need to consider only the effective Hamiltonian described solely by the system operators. Moreover, any additional waveguide couplings via say $a_j$ is dealt with by adding a term $-\i\frac{\gamma_j}{2} a_j^\dagger a_j$ to the effective Hamiltonian. Similarly, any loss to free space through the system operator, $a_j$ is taken care of by adding a term $-\i\frac{\Gamma_j}{2} a_j^\dagger a_j$ to the effective Hamiltonian. 

In this work, we stay in the weak waveguide-local system coupling limit by assuming that $\gamma$ $\ll$ the smallest energy gap of the local system and ignoring the effects of extra decay into free space. All of our results in \Cref{sec:examples} have been worked out within this regime and remain valid as long as the effects of the bandwidth and dispersion of the waveguides can be ignored.

\section{Diagrammatic approach}\label{sec:diagram}
As we discussed, the problem of calculating the $n$-photon S-matrix has been reduced to calculating all $2m$-point Green's functions, for $m \leqslant n$. Here, we aim to provide a quick and straightforward recipe to compute them. As a gentle introduction, let us start by describing how a $2$-point Green's function is calculated. The latter is defined as 
\begin{align*}
G(p; k) &= -\mathscr{F}\Bra{0}\mathcal{T}a(t')a^\dagger(t)\Ket{0} \\*
&= -\mathscr{F}\Big\{\Bra{0}a(t')a^\dagger(t)\Ket{0}\theta(t' - t)\Big\}.
\end{align*}
To proceed, the system operators need to be expressed in the Heisenberg form under the effective Hamiltonian. If the spectrum of the effective Hamiltonian is $H_\text{eff}\Ket{\epsilon} = \epsilon\Ket{\epsilon}$, $\Bra{\bar{\epsilon}}H_\text{eff} = \epsilon\Bra{\bar{\epsilon}}$, with normalisation conditions $\braket{\epsilon|\bar{\epsilon'}} = \delta_{\epsilon\epsilon'}$, where $\epsilon$ denote the eigenenergies and $\Ket{\epsilon}$ and $\Bra{\bar{\epsilon}}$ denote the right and left eigenvectors respectively, the above step gives us
\[G(p; k) = -\mathscr{F}\Big\{\sum_{\epsilon} \braket{0|a|\epsilon}\braket{\bar{\epsilon}|a^\dagger|0}\e^{-\i\epsilon(t' - t)}\theta(t' - t)\Big\}.\]
The sum is taken over all eigenenergies of $H_\text{eff}$. The calculation is completed by performing the Fourier transformation.

Although it is seemingly straightforward to do, it gets cumbersome quickly as the number of photons increases. This motivates us to develop a diagrammatic technique to calculate the quantity which at the same time is able to provide us an intuitive picture of the processes that happen during an inelastic scattering. 
The diagrammatic method to compute the $2m$-point Green's function, $G(p_1\dots p_m; k_1\dots k_m)$ is as follows: (proof is given in \cref{sec:proof})
\begin{enumerate}
\item Draw all possible diagrams in the excitation space where each diagram corresponds to a particular time ordering of the Green's function. All diagrams start and end at the vacuum state (zero excitations) and consist of $m$ input momenta (upwards arrows representing creation of excitation, $a^\dagger$) and $m$ output momenta (downwards arrows representing annihilation of excitation, $a$). Input and output momenta are labeled with $k$'s and $p$'s respectively. \Cref{fig:2point} shows the only possible diagram for the 2-point Green's function while \cref{fig:diagrams} shows all the possible diagrams for 4- and 6-point Green's functions.

\begin{figure}[!h]
\centering
\includegraphics[width=0.3\textwidth]{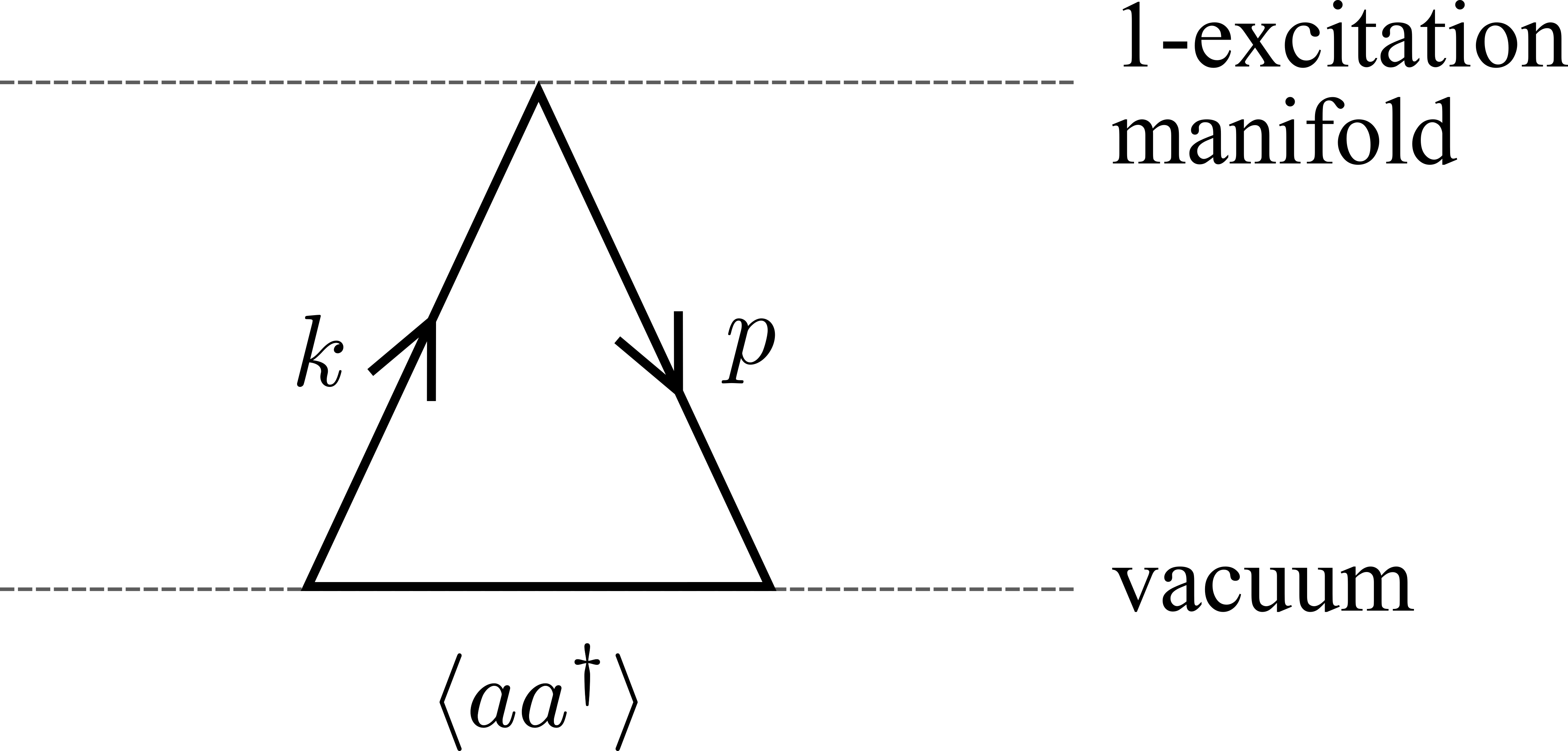}\\
\caption{Diagram for 2-point Green's function.}\label{fig:2point}
\end{figure}

\begin{figure*}[!t]
\centering  
\includegraphics[width=0.8\textwidth]{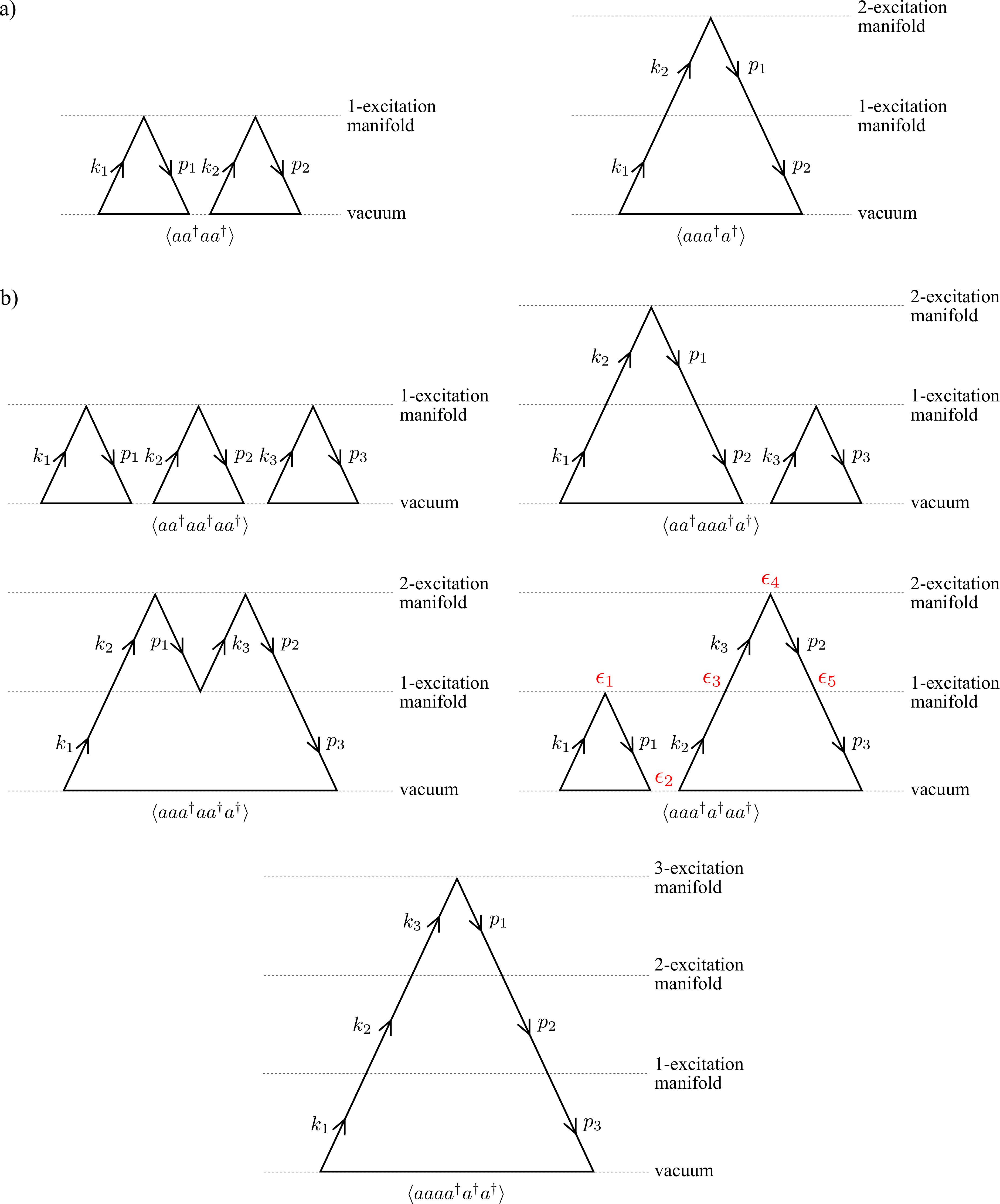}
\caption{Examples of all possible diagrams for a) 4-point Green's funtion and b) 6-point Green's function. Note that some of the diagrams drawn here are simply the disconnected products of the lower order diagrams. Generally, one can construct all diagrams for higher order Green's functions by building upon the lower order diagrams and drawing purely connected ones.}\label{fig:diagrams}
\end{figure*}

\item For each diagram, assign a factor of
\begin{equation*}
\frac{-\i}{(2\pi)^{m - 1}}\;\delta\left(\sum_{i = 1}^m k_i - \sum_{i = 1}^m p_i\right)
\end{equation*}
for momentum conservation. As a consequence of this, at every node between two momenta, the sum of all upwards momenta to the left of the node is equal the sum of all downwards momenta to the right of the node. We label such sums by $K_i$ for $i = 1, \dots, (2m - 1)$. For example, with the diagram $\braket{aaa^\dagger a^\dagger aa^\dagger}$ in \cref{fig:diagrams}(b), we have the sums, $K_1 = k_1 = p_1 + p_2 + p_3 - k_2 - k_3$, $K_2 = k_1 - p_1 = p_2 + p_3 - k_2 - k_3$, $K_3 = k_1 + k_2 - p_1 = p_2 + p_3 - k_3$, $K_4 = k_1 + k_2 + k_3 - p_1 = p_2 + p_3$ and $K_5 = k_1 + k_2 + k_3 - p_1 - p_2 = p_3$.

\item Next, assign a `propagator' for each arrow except the last one. It is written as
\begin{equation*}
\prod_{j = 1}^{2m - 1}\frac{1}{K_j - \epsilon_j},
\end{equation*}
where $\epsilon_j$'s are dummy variables for the eigenenergies of $H_\text{eff}$ that will be summed over in the next step. 

\item Finally, sum over all the eigenenergies with the respective weights
\begin{equation*}
\sum_{\epsilon_1, \dots, \epsilon_{2m - 1}}\braket{0|\cdot|\epsilon_{2m - 1}}\braket{\bar{\epsilon}_{2m - 1}|\dots|\epsilon_1}\braket{\bar{\epsilon}_1|\cdot|0}
\end{equation*}
and over all permutations of input and output momenta which leave the diagram unchanged. For example, in \cref{fig:diagrams}(b), the Green's function corresponding to the diagram $\braket{aaa^\dagger a^\dagger aa^\dagger}$ yields
\begin{align*}
&-\frac{\i}{4\pi^2}\delta(k_1 + k_2 + k_3 - p_1 - p_2 - p_3)\cdot \\*
\sum_{\epsilon_1, \epsilon_3, \epsilon_4, \epsilon_5} &\braket{0|a|\epsilon_5}\braket{\bar{\epsilon}_5|a|\epsilon_4}\braket{\bar{\epsilon}_4|a^\dagger|\epsilon_3}\braket{\bar{\epsilon}_3|a^\dagger|0} \\*
&\braket{0|a|\epsilon_1}\braket{\bar{\epsilon}_1|a^\dagger|0}\frac{1}{k_1 - \epsilon_1}\frac{1}{k_1 - p_1} \\*
&\frac{1}{k_1 + k_2 - p_1 - \epsilon_3}\frac{1}{p_2 + p_3 - \epsilon_4}\frac{1}{p3 - \epsilon_5} \\*
\text{+ all\;} &\text{permutations of $\{k_1, k_2, k_3\}$ and $\{p_1, p_2, p_3\}$}
\end{align*}
where the summation is over all eigenenergies for $\epsilon_1$, $\epsilon_3$, $\epsilon_4$ and $\epsilon_5$. Since the ground state is unique and is always set to have zero energy, $\epsilon_2$ (which has been left out in the expression) is zero.
\end{enumerate}

Following these four steps, one can write down the connected parts of the Green's function (which is also the connected parts of the S-matrix) in a systematic way. Once they are found, the cluster decomposition structure illustrated in \cref{eq:clusterd} and \cref{sec:clusterS} is used to calculate the full S-matrix. 

\section{Paradigmatic examples}\label{sec:examples}
In this section, several examples are presented to demonstrate the diagrammatic approach detailed above. The examples consist of systems that are total-particle-number conserving, $[H_\text{sys}, N_\text{tot}] = 0$, starting from non-interacting systems, $H_\text{int} = 0$ with a two-level emitter and two collocated two-level emitters and ending with a many-emitter interacting system described by the Bose-Hubbard model. Implementations of such systems have been explored in areas such as superconducting circuits \citep{wallraff04}, cold atoms \citep{bloch12}, and cavity systems \citep{exploring}. 
 
\subsection{Two-level system}
\begin{figure}[h]
  \centering
  \includegraphics[width=0.35\textwidth]{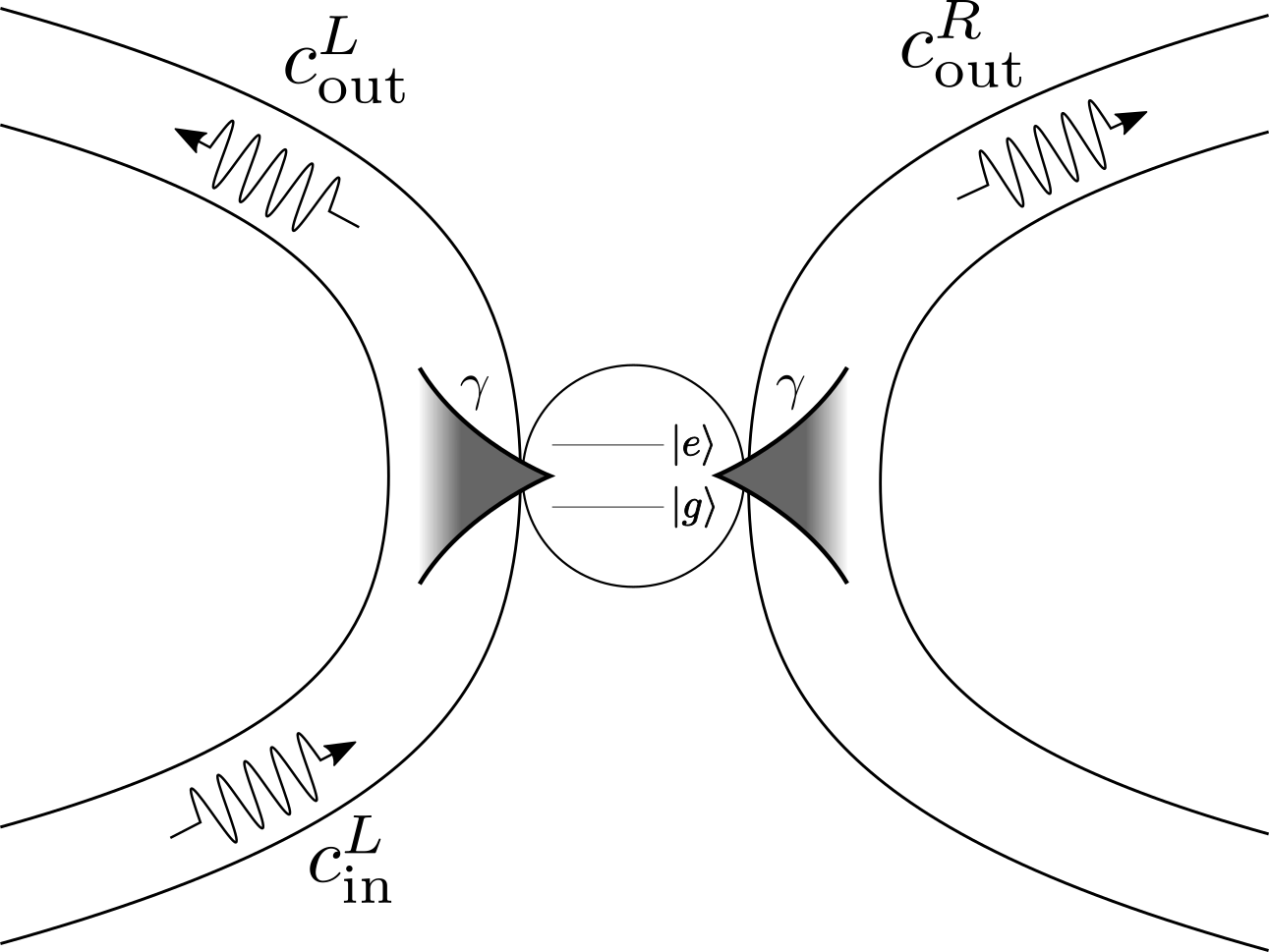}\\
  \caption{Scattering photons from a two-level system.}\label{fig:2level}
\end{figure}

First, we consider a simple example --- a two-level system described by $H_\text{sys} = H_\text{2-lvl} = \omega\sigma_{ee}$ where $\sigma_{ee} = \ket{e}\bra{e}$, embedded between two waveguides (\cref{fig:2level}). The effective Hamiltonian is simply given by
\[H_\text{eff} = (\omega - \i\gamma)\sigma_{ee}.\]

\begin{figure}[h]
  \centering
  \includegraphics[width=0.48\textwidth]{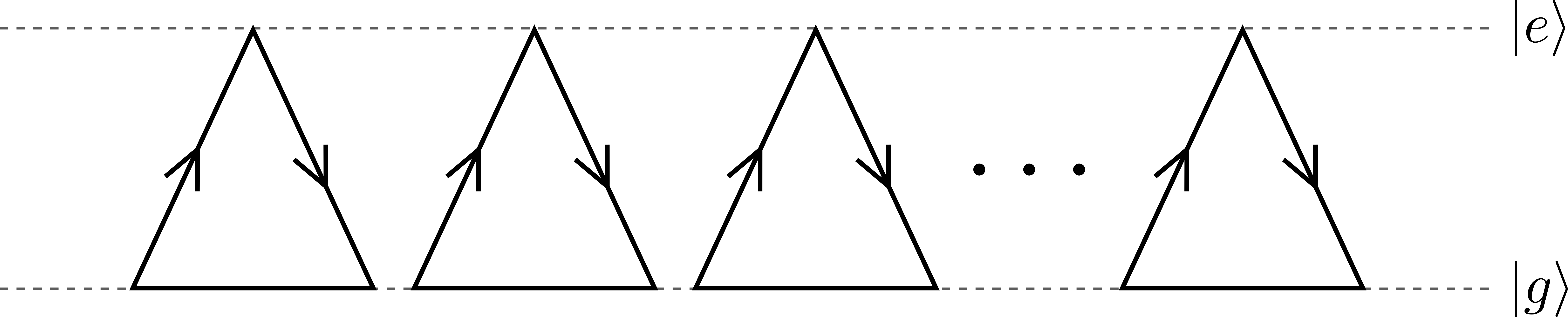}\\
  \caption{The only possible scattering diagram of the $2m$-point Green's functions for a two level quantum emitter being probed by $m$ photons.}\label{fig:2diag}
\end{figure}

Following the first step in \cref{sec:diagram}, we start by drawing all possible diagrams for Green's functions. Since the effective Hamiltonian is diagonal, comprised of only the ground state $\ket{g}$ and a single excitation state $\ket{e}$, only possible diagrams are the ones that are products of the triangular loop diagram of the 2-point Green's function (\cref{fig:2diag}). The corresponding $2m$-point Green's function is
\begin{align}\label{eq:2levelG}
G_{\vec{i};\vec{j}} (\vec{p}; \vec{k}) = &- \frac{\i\gamma^m}{(2\pi)^{m - 1}}\delta(\vec{p} - \vec{k})\Bigg(\frac{1}{k_1 - \omega + \i\gamma}\frac{1}{k_1 - p_1}\cdot \nonumber\\*
&\frac{1}{k_1 + k_2 - p_1 - \omega + \i\gamma}\cdots\frac{1}{p_m - \omega + \i\gamma} \nonumber\\*
&+ \text{permutations of $\{\vec{p}\}$ and $\{\vec{k}\}$}\Bigg).
\end{align}
$\vec{i}$ and $\vec{j}$ are the labels $L$/$R$, for the outgoing and incoming photon channels with corresponding outgoing momenta $\vec{p}$ and incoming momenta $\vec{k}$, respectively. Note that the expressions of the Green's functions for different sets of outgoing and incoming photon channels are identical and are distinguished by the corresponding $\vec{i}$ and $\vec{j}$ labels associated to the momenta. 
Furthermore, inserting photons only from the left channel gives no extra delta function in the one-photon S-matrix of the right channel output, because there is no event where a photon would pass by the right channel without going through the two-level atom.

Following the rest of the steps in \cref{sec:diagram}, we obtain the one-photon S-matrices
\begin{equation}\label{eq:oneS}
\begin{array}{rcccl}
S_{L;L}(p; k) &=& \delta(p - k) + G_{L;L}(p;k) &=& r_k\delta(p - k), \\*
S_{R;L}(p; k) &=& G_{R;L}(p;k) &=& t_k\delta(p - k),
\end{array}
\end{equation}
where
\begin{align*}
r_k = 1 -\i\gamma\frac{1}{k - \omega + \i\gamma}, \;\;\;\;\;\;\; t_k = -\i\gamma\frac{1}{k - \omega + \i\gamma},
\end{align*}
are the reflection and transmission coefficients. Using the cluster decomposition relation, one can derive the two-photon S-matrices
\begin{align}\label{eq:2S}
S_{ij;LL}(p_1, p_2;k_1, k_2) &= G_{ij;LL}(p_1, p_2;k_1, k_2) + \nonumber\\*
&\hspace{-0.5cm}\left(S_{i;L}(p_1;k_1)S_{j;L}(p_2;k_2) + (k_1 \leftrightarrow k_2)\right)
\end{align}
with the 4-point Green's functions defined in \cref{eq:2levelG} and have expressions
\begin{align}
&G_{ij;LL}(p_1,p_2;k_1,k_2) \nonumber\\*
= &\frac{\i\gamma^2(k_1 + k_2 - 2\omega + 2\i\gamma)}{\pi(k_1 - \omega + \i\gamma)(k_2 - \omega + \i\gamma)(p_1 - \omega + \i\gamma)(p_2 - \omega + \i\gamma)}\cdot \nonumber\\*
&\delta(p_1 + p_2 - k_1 - k_2)
\end{align}
where $i$ and $j$ are the labels $L$/$R$ for outgoing photon channels with output momenta $p_1$ and $p_2$ respectively. Within our diagrammatic approach, the aforementioned 4-point Green's function is visualised in \cref{fig:2diag} but with two loops only. Higher-order S-matrices can be written down similarly using the cluster decomposition relation with the Green's functions in \cref{eq:2levelG}.

To see the significance of this result, consider first a linear system (for example, a simple harmonic oscillator). Such systems have all the connected $2m$-point Green's functions being identically zero for $m \geq 2$ and hence the $n$-photon S-matrices are just products of the one-photon S-matrix for $n \geq 2$. For the case where $n$ photons come in from the left and exits to the right, the $n$-photon S-matrix is $S_{\vec{p};\vec{k}} \propto t_{k_1}t_{k_2}\dots t_{k_n}$. In particular, if the input field is coherent, $S_{\vec{p};\vec{k}} \propto (t_k)^n$, yielding a coherent output field. Therefore, in order to obtain a non-classical output, some kind of nonlinearity in the system is a must. In this example, the two-level atom is strongly nonlinear. Hence, the output is modified from being classical-like by the presence of nonzero Green's functions. Moreover, the diagrams of the Green's functions (\cref{fig:2diag}) match the description of the photon blockade effect where photons are being absorbed and emitted one at a time \citep{birnbaum05,birnbaum052}. Of course one could then proceed with the calculation of the transmitted light correlation functions using the S-matrix derived above to reproduce the anti-bunching correlation/photon blockade effect well known to this system. We would like to point out here that such an effect is intuitive and obvious from the diagrams of the Green's functions (see \cref{fig:2diag}). Calculation of the S-matrix using other methods without these diagrams would not give us such an insight. 

\subsection{Two collocated atoms}\label{sec:2coll}
\begin{figure}[!h]
  \centering
  \includegraphics[width=0.48\textwidth]{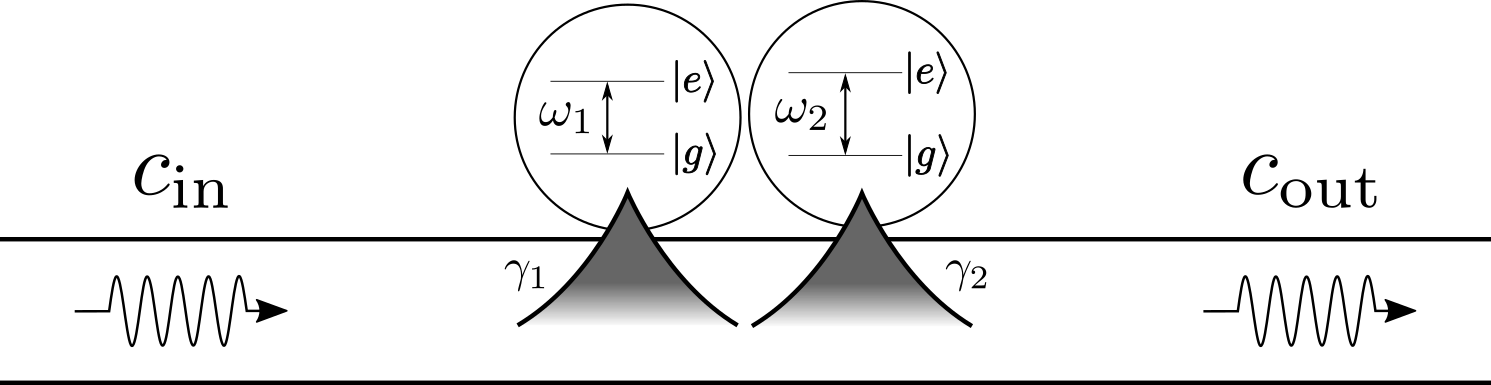}
  \caption{Schematic diagram of two collocated (ignore the distance in the sketch) two-level atoms coupled to a waveguide.}\label{fig:twocoll}
\end{figure}

Before moving into a many-correlated-emitter example, we step up the complexity of the previous example by looking at two collocated non-interacting two-level atoms \citep{rephaeli11}, with $H_\text{sys} = H_\text{2-coll} = \omega_1\sigma_{ee}^{(1)} + \omega_2\sigma_{ee}^{(2)}$. This example, in the limiting case of two identical atoms and coupling strengths, describes the two-atom Dicke model \citep{garraway11}. Working with only one channel that both atoms are coupled to (\cref{fig:twocoll}), we have the total Hamiltonian
\begin{align*}
H_\text{tot} &= H_\text{2-coll} + \int\dd k\; kc_k^{\dagger} c_k^{\vphantom{\dagger}} \\ 
&\quad + \int\dd k\; \left(c^\dagger_k(\xi^{\vphantom{\dagger}}_1\sigma_{ge}^{(1)} + \xi^{\vphantom{\dagger}}_2\sigma_{ge}^{(2)}) + h.c.\right) ,
\end{align*}
where $\xi_{i}$ is the coupling strength between the the $i$th atom and the waveguide modes, for $i = 1, 2$. This gives rise to an input-output relation
\begin{equation}\label{eq:2collinout}
c_\text{out}(t) = c_\text{in}(t) - \i\sqrt{\gamma_1}\sigma_{ge}^{(1)}(t) - \i\sqrt{\gamma_2}\sigma_{ge}^{(2)}(t).
\end{equation}

Comparing the forms of \cref{eq:2collinout,eq:inout}, we see that the $\sqrt{\gamma}a$ in \cref{eq:inout} is replaced by $\sqrt{\gamma_1}\sigma_{ge}^{(1)} + \sqrt{\gamma_2}\sigma_{ge}^{(2)}$. Therefore, the effective Hamiltonian can be deduced and takes the form:
\begin{align*}
H_\text{eff} &= \omega_1\sigma_{ee}^{(1)} + \omega_2\sigma_{ee}^{(2)} \\*
&\quad - \frac{\i}{2}(\sqrt{\gamma_1}\sigma_{ge}^{(1)} + \sqrt{\gamma_2}\sigma_{ge}^{(2)})^\dagger(\sqrt{\gamma_1}\sigma_{ge}^{(1)} + \sqrt{\gamma_2}\sigma_{ge}^{(2)}) \\*
&= (\omega_1 - \i\frac{\gamma_1}{2})\sigma_{ee}^{(1)} + (\omega_2 - \i\frac{\gamma_2}{2})\sigma_{ee}^{(2)} \\*
&\quad - \i\frac{\sqrt{\gamma_1\gamma_2}}{2}(\sigma_{eg}^{(1)}\sigma_{ge}^{(2)} + \sigma_{ge}^{(1)}\sigma_{eg}^{(2)}) 
\end{align*}

\begin{figure*}[!th]
  \centering
  \includegraphics[width=1\textwidth]{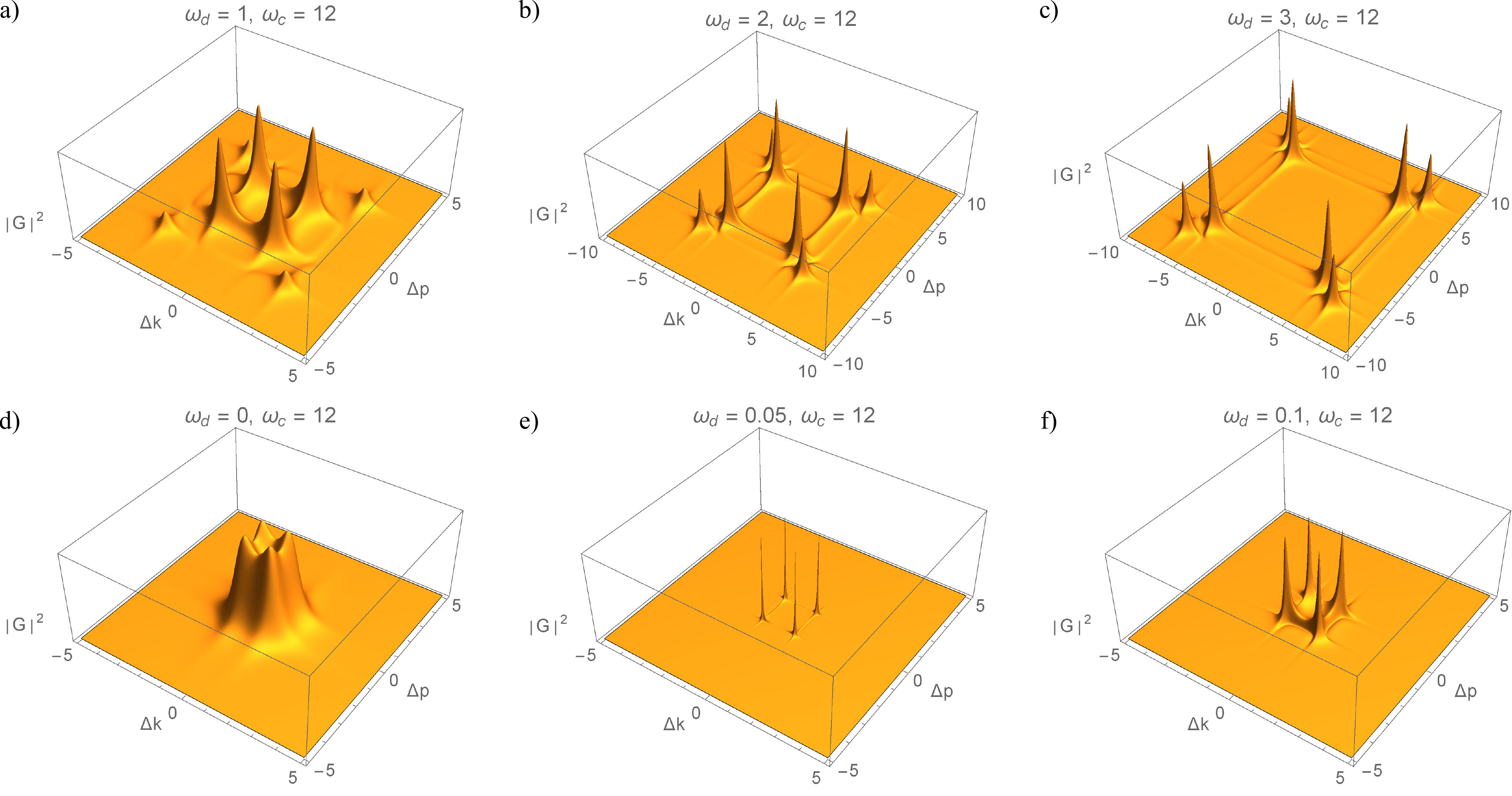}
  \caption{\textbf{Two collocated atoms:} Plots of $|G(p_1,p_2;k_1,k_2)|^2$ where i) $\omega_d^2 > \gamma^2_c/4$ with $\omega_d = 1, 2$ and $3$ (figures (a--c)); and ii) $\omega_d^2 < \gamma^2_c/4$ with $\omega_d = 0, 0.05$ and $0.1$ (figures (d--f)). For all the plots, $\omega_c = 12$, $\gamma_c = 0.25$ and the total incoming momentum, $E_i = 2\omega_c + 3\gamma_c$. We define $\Delta k \equiv k_1 - k_2$ and $\Delta p\equiv p_1 - p_2$. In figures (a--c), under i), the spectra exhibit eight peaks with the same width while in figures (d--f), under ii), the spectra exhibit only four peaks with the width becoming smaller as the detuning, $\omega_d$ decreases except at zero detuning where the width is much wider.}\label{fig:2coll}
\end{figure*}

The effective Hamiltonian shows that even though the two collocated atoms do not interact directly, coupling to the same waveguide induces an effective dissipative interaction between them. Moreover, it has a spectrum: $\epsilon^{(0)} = 0, \epsilon^{(1)}_{\pm} = \omega_c - \i\gamma_c/2 \pm \sqrt{\omega_d^2 - \i\omega_d\gamma_d - \gamma_c^2/4}, \epsilon^{(2)} = 2\omega_c - \i\gamma_c$, where $\omega_c = \frac{\omega_1 + \omega_2}{2},\; \omega_d = \frac{\omega_1 - \omega_2}{2},\; \gamma_c = \frac{\gamma_1 + \gamma_2}{2}$, and $\gamma_d = \frac{\gamma_1 - \gamma_2}{2}$. The corresponding right eigenvectors are $\ket{\epsilon^{(0)}} = \ket{gg}$, $\ket{\epsilon^{(1)}_\pm} \propto \i\sqrt{\gamma_c^2 - \gamma_d^2}/4\ket{eg} + (\omega_d - \i\gamma_d/2 \mp \sqrt{\omega_d^2 - \i\omega_d\gamma_d - \gamma_c^2/4})\ket{ge}$ and $\ket{\epsilon^{(2)}} = \ket{ee}$. The left eigenvectors have the same coefficients but with the corresponding $\bra{\cdot}$. The single excitation manifold, $\epsilon_\pm^{(1)}$ becomes degenerate when 
\begin{equation}\label{eq:2cond}
\gamma_d = 0,\; \omega_d^2 = \gamma_c^2/4
\end{equation}
and the effective Hamiltonian becomes undiagonalisable. Hence, the theory cannot be applied under this condition. However, we can still study how the system behaves when $\gamma_d = 0$ (i.e. $\gamma_1 = \gamma_2 = \gamma_c$), but $\omega_d^2 \neq \gamma_c^2/4$. We noticed that when, i) $\omega_d^2 > \gamma^2_c/4$, $\epsilon^{(1)}_{\pm} = \omega_{1, 2} - \i\gamma_c/2 + \mathcal{O}(\frac{\gamma_c^2}{\omega_d^2})$; and when ii) $\omega_d^2 < \gamma^2_c/4$, 
\[\epsilon^{(1)}_{\pm} = \left\{\begin{array}{l}
\omega_c - \i\omega_d^2/\gamma_c + \mathcal{O}(\frac{\omega_d^4}{\gamma_c^4}) \\
\omega_c - \i\gamma_c + \mathcal{O}(\frac{\omega_d^2}{\gamma_c^2})
\end{array}\right..\]

Since the peaks we observe in the S-matrix are closely related to the spectrum of the effective Hamiltonian, the condition in \cref{eq:2cond} defines some sort of critical point separating the region where i) the atoms decay as independent atoms and ii) the atoms decay collectively either with a subradiant (radiation reduction) or superradiant (radiation enhancement) profile. Moreover, in ii) when detuning is zero, $\omega_d = 0$, the superradiant state, $\ket{\epsilon^{(1)}_-} = (\ket{eg} + \ket{ge})/\sqrt{2}$, that decays as $\gamma_c$ is the only state that is coupled to the input field. To visualise the two different situations, we first write the 2-point Green's function and one-photon S-matrix,
\begin{align*}
G(p;k) &= \sum_{\epsilon^{(1)}}\braket{0|a|\epsilon^{(1)}}\braket{\bar{\epsilon}^{(1)}|a^\dagger|0}\frac{-\i}{k - \epsilon^{(1)}}\delta(k - p) \\*
&= \frac{-2\i\gamma_c(k - \omega_c)}{(k - \omega_1 + \frac{\i\gamma_c}{2})(k - \omega_2 + \frac{\i\gamma_c}{2}) + \frac{\gamma_c^2}{4}}\delta(k - p) \\*
&\equiv g_k \delta(k - p)
\end{align*}
and
\begin{align*}
S(p;k) &= \delta(k - p) + G(p;k) \\*
&= \frac{(k - \omega_1 - \frac{\i\gamma_c}{2})(k - \omega_2 - \frac{\i\gamma_c}{2}) + \frac{\gamma_c^2}{4}}{(k - \omega_1 + \frac{\i\gamma_c}{2})(k - \omega_2 + \frac{\i\gamma_c}{2}) + \frac{\gamma_c^2}{4}}\delta(k - p). 
\end{align*}

Next, the 4-point Green's function can be computed following the steps in \cref{sec:diagram} with the aid of the diagrams in \cref{fig:diagrams}.
\begin{align}\label{eq:1_2g}
&G(p_1, p_2; k_1, k_2) \nonumber\\*
= &\sum_{l = 1}^2 G^{(l)}(p_1, p_2; k_1, k_2)\delta(E_o - E_i)
\end{align}
where $E_i = k_1 + k_2$, $E_o = p_1 + p_2$, and $G^{(1)}$, $G^{(2)}$ represent the diagrams $\braket{aa^\dagger aa^\dagger}$, $\braket{aaa^\dagger a^\dagger}$ respectively, with expressions
\begin{align}\label{eq:1_2g12}
&G^{(1)}(p_1, p_2; k_1, k_2) \nonumber\\*
= &-\frac{\i}{2\pi}\sum_{\epsilon^{(1)}_1, \epsilon^{(1)}_2}\Bigg(\frac{\braket{0|a|\epsilon^{(1)}_2}\braket{\bar{\epsilon}^{(1)}_2|a^\dagger|0}}{p_2 - \epsilon^{(1)}_2}\frac{1}{k^{\vphantom{(1)}}_1 - p^{\vphantom{(1)}}_1}\cdot \nonumber\\*
&\hspace{10em}\frac{\braket{0|a|\epsilon^{(1)}_1}\braket{\bar{\epsilon}^{(1)}_1|a^\dagger|0}}{k_1 - \epsilon^{(1)}_1}\Bigg) \nonumber\\*
&+ \;\text{all permutations of $\{k_1, k_2\}$ and $\{p_1, p_2\}$} \\
\intertext{and}
&G^{(2)}(p_1, p_2; k_1, k_2) \nonumber\\*
= &-\frac{\i}{2\pi}\sum_{\epsilon^{(1)}_1, \epsilon^{(2)}_2, \epsilon^{(1)}_3}\Bigg(\braket{0|a|\epsilon^{(1)}_3}\braket{\bar{\epsilon}^{(1)}_3|a|\epsilon^{(2)}_2}\braket{\bar{\epsilon}^{(2)}_2|a^\dagger|\epsilon^{(1)}_1}\cdot \nonumber \\*
&\hspace{3em}\braket{\bar{\epsilon}^{(1)}_1|a^\dagger|0}\frac{1}{k_1 - \epsilon^{(1)}_1}\frac{1}{k_1 + k_2 - \epsilon^{(2)}_2}\frac{1}{p_2 - \epsilon^{(1)}_3}\Bigg)\nonumber\\
&+ \;\text{all permutations of $\{k_1, k_2\}$ and $\{p_1, p_2\}$}.
\end{align}

The exact expression for the 4-point Green's function is listed in \cref{sec:exact2coll}. Finally, the two-photon S-matrix is given by
\begin{align}\label{eq:1_2S}
S(p_1, p_2;k_1, k_2) &= G(p_1, p_2;k_1, k_2) + \nonumber\\*
&\hspace{-0.5cm}\left(S(p_1;k_1)S(p_2;k_2) + (k_1 \leftrightarrow k_2)\right). 
\end{align}

Graphs of $|G(p_1, p_2;k_1, k_2)|^2$ are plotted in \cref{fig:2coll} under two different conditions i) $\omega_d^2 > \gamma^2_c/4$ and ii) $\omega_d^2 < \gamma^2_c/4$. The plots are generated with the total incoming momentum going slightly off the two-photon resonance ($E_i = 2\omega_c$) because the 4-point Green's function is identically zero at resonance (\cref{sec:exact2coll}). This means that two photons inserted into the system at the two-photon resonance will scatter off as independent photons. In \cref{fig:2coll}(a--c), under i), the spectra exhibit eight peaks with the same width while in \cref{fig:2coll}(d--f), under ii), the spectra exhibit only four peaks and the width of the peaks becomes smaller as the detuning $\omega_d$ decreases except at zero detuning where the peaks are wider. This observation is consistent with the above arguments that under i) the atoms decay like independent particles, hence the spectra exhibit eight peaks and with the same decay rate which is represented by the width of the peaks while under ii) the atoms decay collectively hence the spectra exhibit only half the number of peaks as compared to i) and with a dominant subradiant profile (narrow peaks) except at $\omega_d = 0$ where only the superradiant state is coupled, and the superradiant effect can be observed (wide peaks).

\subsection{Bose-Hubbard model}
Finally, we consider a fully interacting, many-body scenario --- the Bose-Hubbard model,
\begin{align*}
H_\text{sys} = H_\text{bh} = &\sum_{j = 1}^N \left(\omega_j a^\dagger_j a_j + \frac{U_j}{2} a^\dagger_j a^\dagger_j a_j a_j\right)\\* 
&+ J\sum_{j = 1}^{N - 1}\left(a^\dagger_j a_{j + 1} + a^\dagger_{j + 1} a_j\right)
\end{align*}
The Bose-Hubbard model has been realised in experiments with ultracold atoms \citep{greiner02} and there are many proposals to realise it in different platforms, notably the one using coupled cavity arrays \citep{angelakis07,hartmann06,greentree06}. This model is interesting as it exhibits rich quantum phases. We study the model assuming two input photons and a mesoscopic number of sites as studied in Refs. \citep{noh16,fitzpatrick17}.

\begin{figure}[!h]
  \centering
  \includegraphics[width=0.48\textwidth]{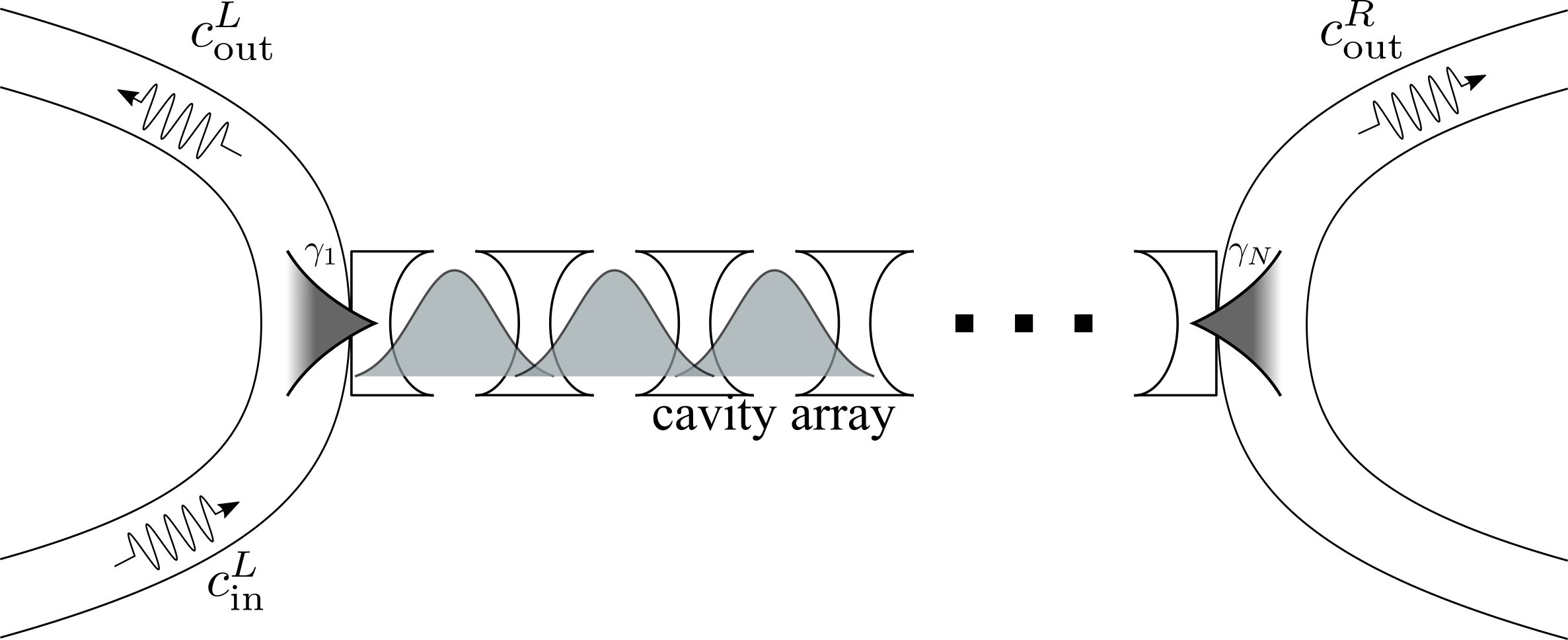}
  \caption{Sketch of a one-dimensional nonlinear cavity array coupled to input and output waveguides.}\label{fig:bhm}
\end{figure}

\Cref{fig:bhm} illustrates a realisation of the Bose-Hubbard model with the open boundary condition, $H_\text{bh}$ coupled to two waveguides via operators $a_1$ and $a_N$. The effective Hamiltonian can be written as 
\[H_\text{eff} = H_\text{bh} - \i\frac{\gamma_1}{2}a_1^\dagger a_1 - \i\frac{\gamma_N}{2}a_N^\dagger a_N.\]

\begin{figure*}[!ht]
  \centering
  \includegraphics[width=0.32\textwidth]{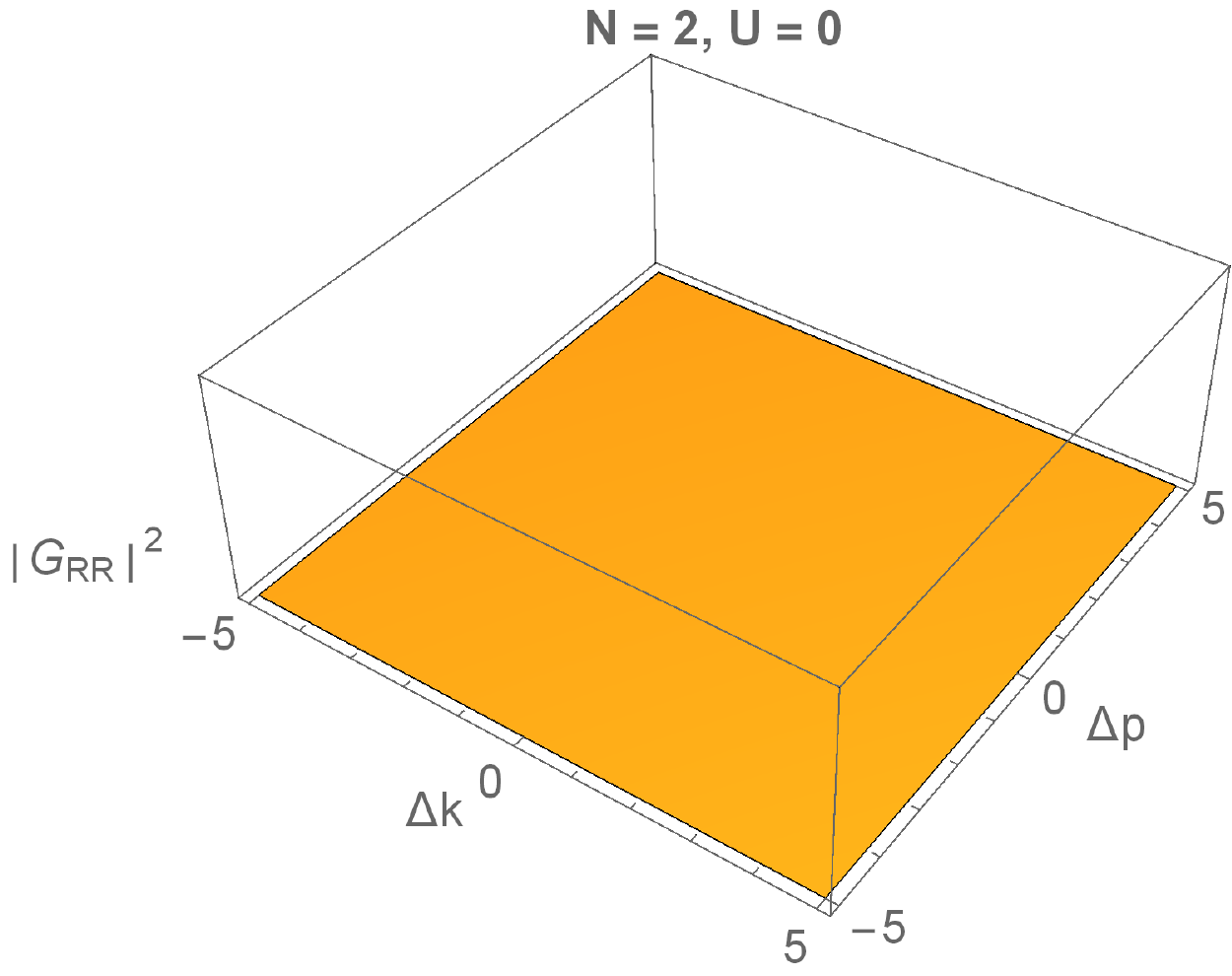}
  \includegraphics[width=0.32\textwidth]{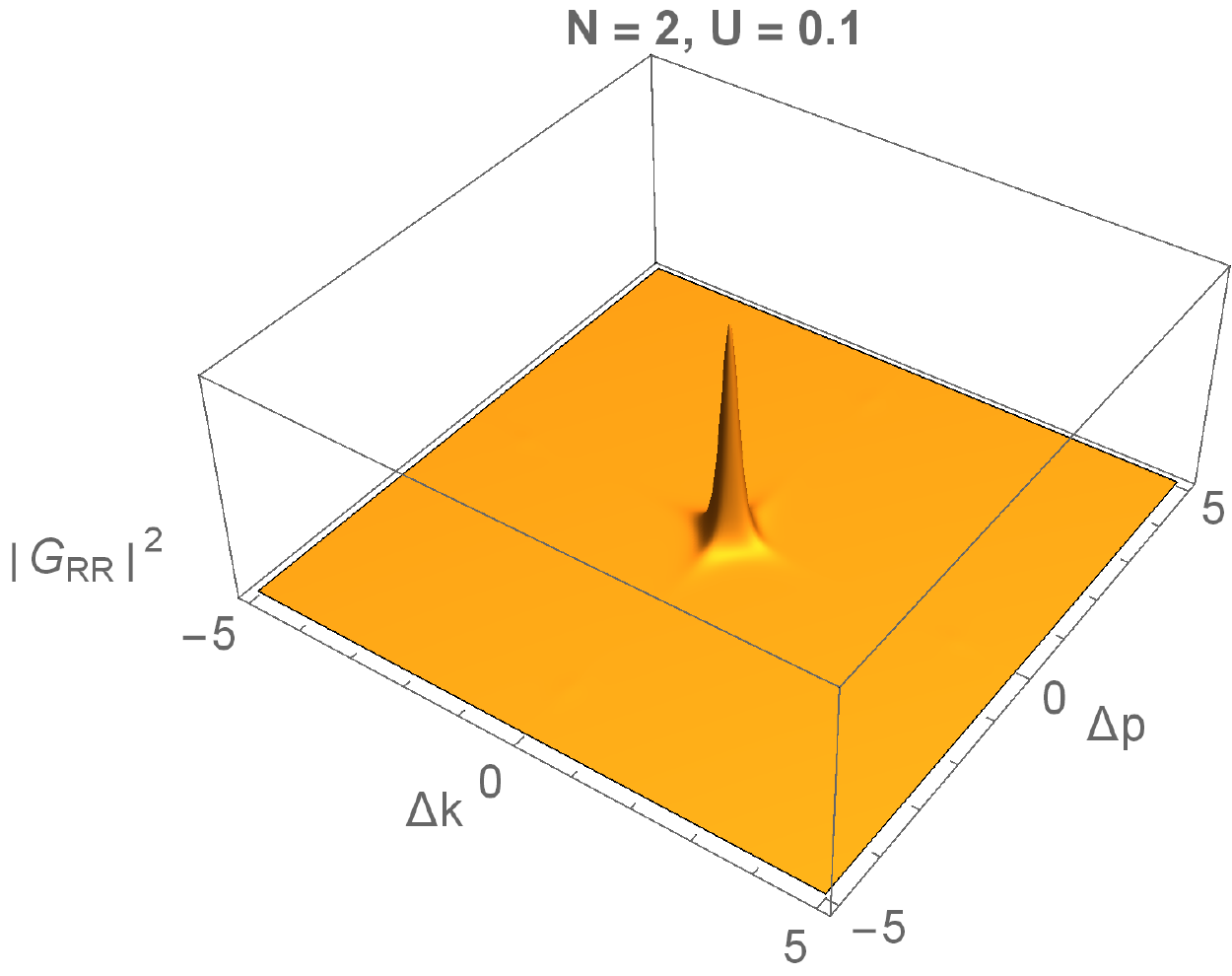}
  \includegraphics[width=0.32\textwidth]{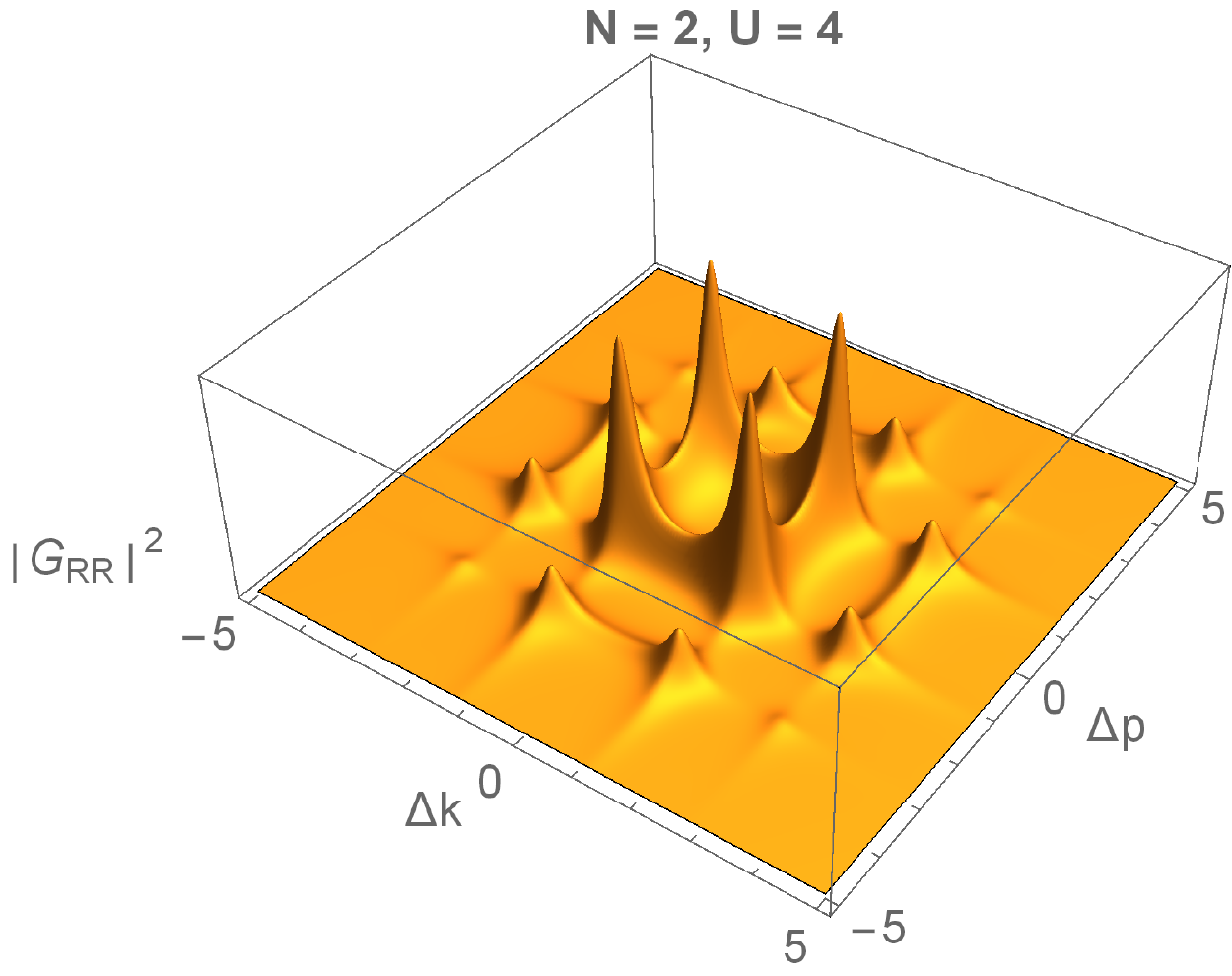}\\
  \includegraphics[width=0.32\textwidth]{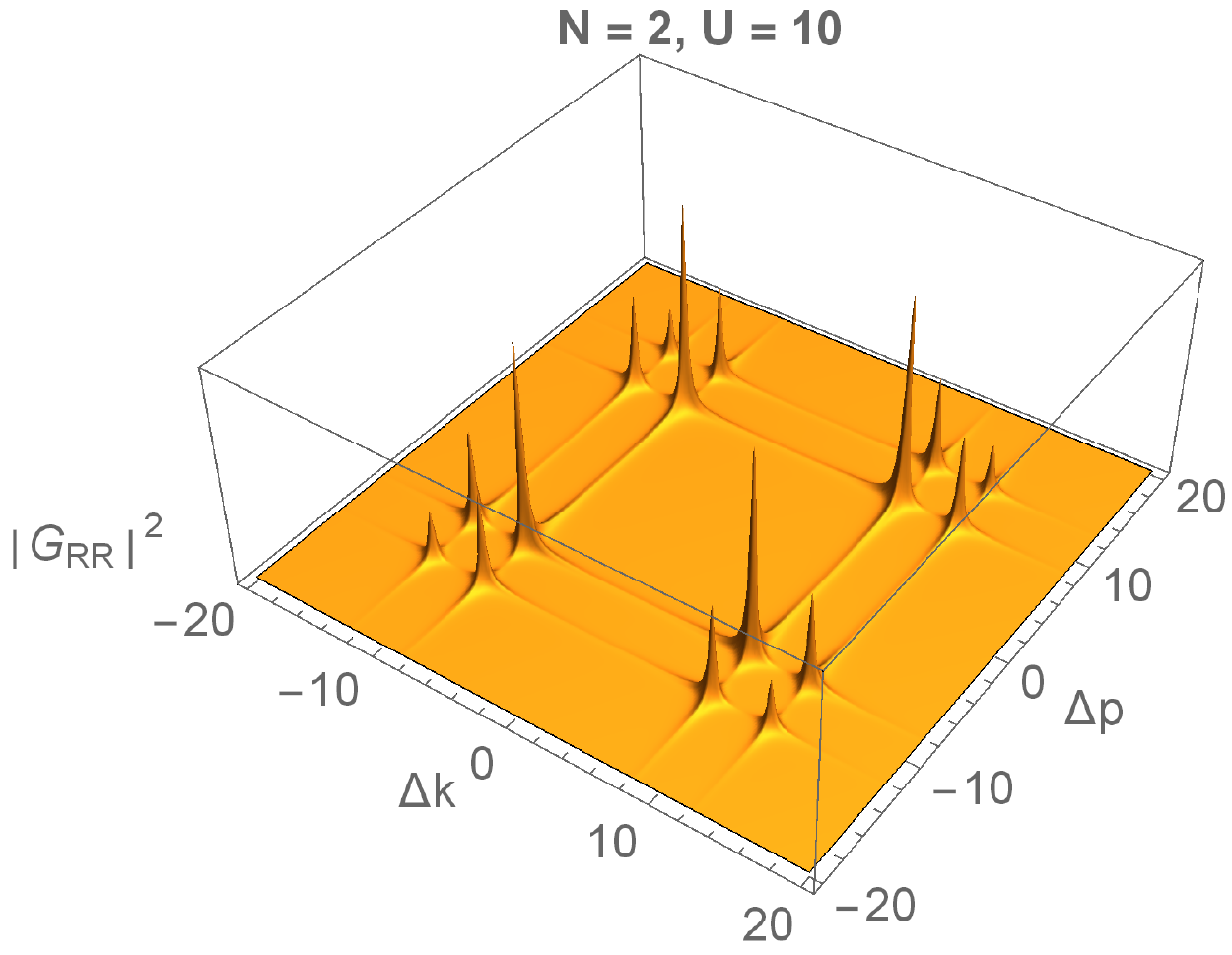}
  \includegraphics[width=0.32\textwidth]{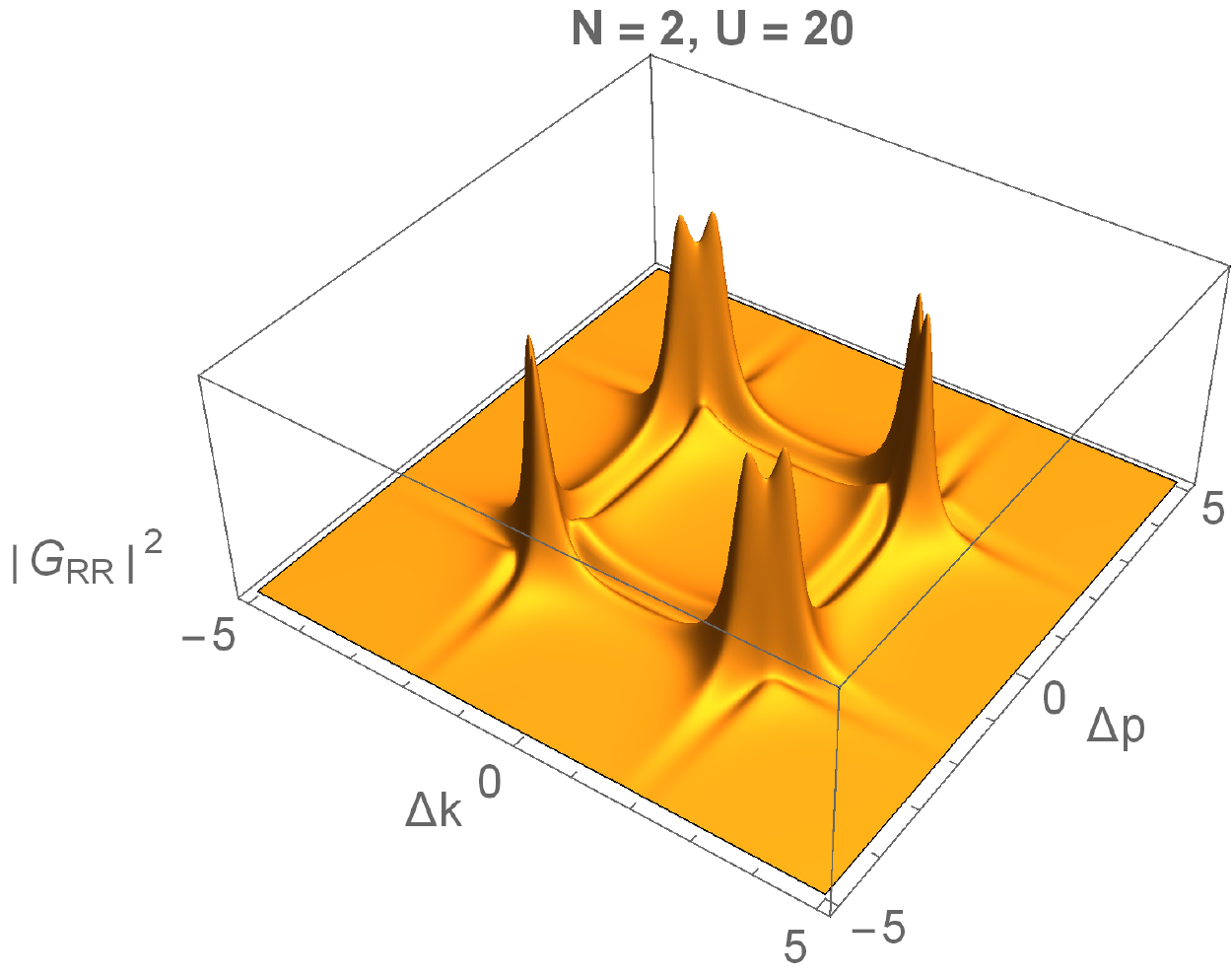}
  \includegraphics[width=0.32\textwidth]{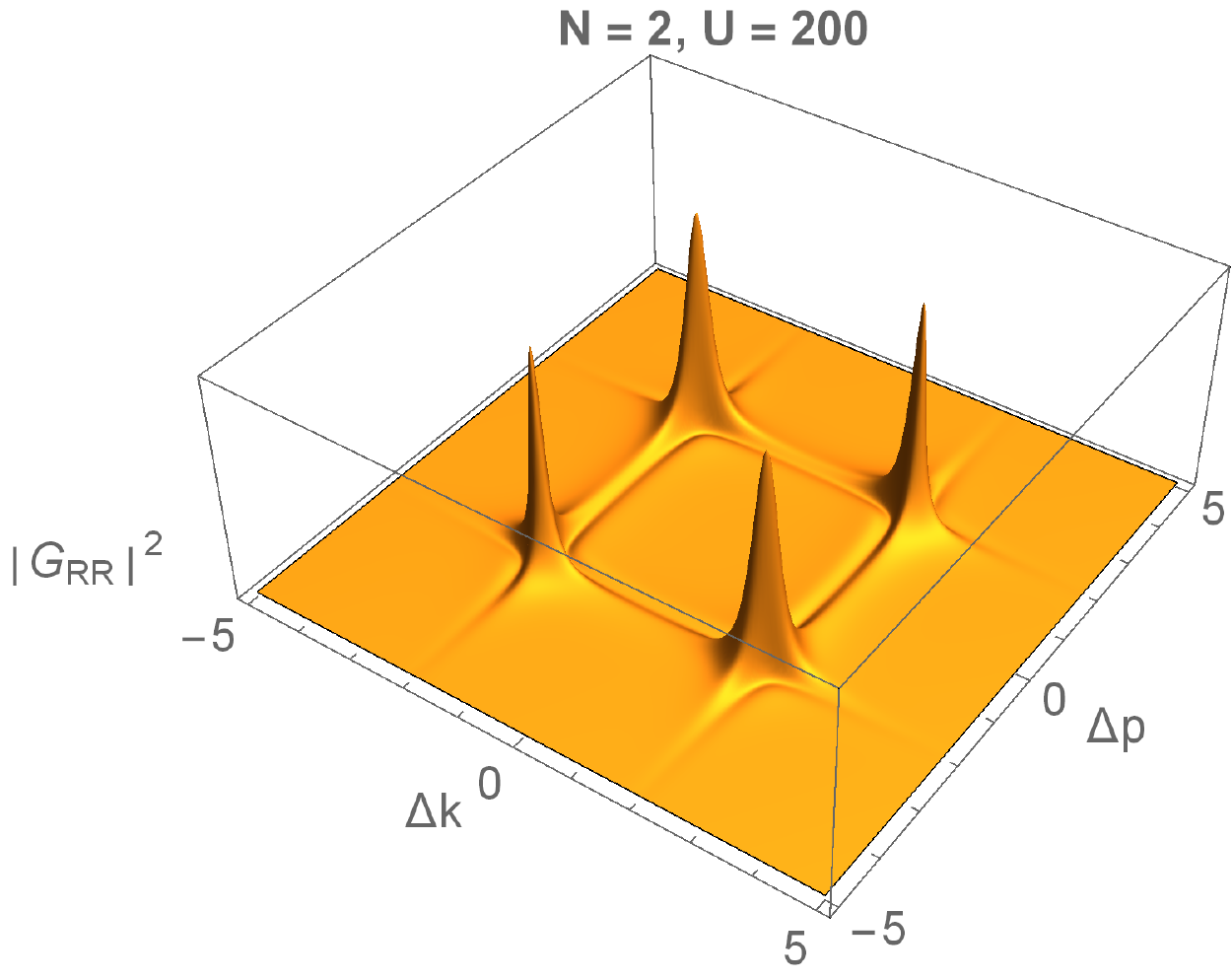}
  \caption{\textbf{Bose-Hubbard model with $N = 2$:} Plots of $|G_{RR;LL}|^2$ (incoming photon channel labels, $LL$ are dropped) of Bose-Hubbard model with $N = 2$. $U = 0, 0.1, 4, 10, 20$ and $200$ respectively above. $\Delta k \equiv k_1 - k_2$ and $\Delta p\equiv p_1 - p_2$. $\omega_0 = 100$, $\gamma = 0.25$ and the total incoming momentum, $k_1 + k_2 = 2\omega_0 + \frac{U}{2} - \sqrt{4J^2 + \frac{U^2}{4}}$, all in units of $J$.}\label{fig:dimer}
\end{figure*}

As before we consider only inputs from the left waveguide. To calculate the one- and two-photon S-matrix, we first compute the 2- and 4-point Green's functions. Following the steps in \cref{sec:diagram} with \cref{fig:2point}, the 2-point Green's functions are
\begin{align}\label{eq:g1bh}
G_{L;L}(p;k) &= \sum_{\epsilon^{(1)}}\braket{0|a_1|\epsilon^{(1)}}\braket{\bar{\epsilon}^{(1)}|a^\dagger_1|0}\frac{-\i\gamma_1}{k - \epsilon^{(1)}}\delta(p - k) \nonumber\\
G_{R;L}(p;k) &= \sum_{\epsilon^{(1)}}\braket{0|a_N|\epsilon^{(1)}}\braket{\bar{\epsilon}^{(1)}|a^\dagger_1|0}\frac{-\i\sqrt{\gamma_1\gamma_N}}{k - \epsilon^{(1)}}\delta(p - k)
\end{align}
where the superscript on $\epsilon$ denotes the particle-number manifold we are summing over. The one-photon S-matrix elements are as shown in \cref{eq:oneS} with the 2-point Green's functions in \cref{eq:g1bh}. Next, from \cref{fig:diagrams}(a), we have
\begin{align}\label{eq:2green}
&G_{i_1 i_2;LL}(p_1, p_2; k_1, k_2) \nonumber\\*
= &\sum_{l = 1}^2 G^{(l)}_{i_1 i_2;LL}(p_1, p_2; k_1, k_2)\delta(p_1 + p_2 - k_1 - k_2)
\end{align}
where $G^{(1)}$ and $G^{(2)}$ represent the diagrams $\braket{aa^\dagger aa^\dagger}$ and $\braket{aaa^\dagger a^\dagger}$ respectively and have expressions
\begin{align}\label{eq:2green12}
&G^{(1)}_{i_1 i_2;LL}(p_1, p_2; k_1, k_2) \nonumber\\*
= &\frac{-\i\gamma_1\sqrt{\gamma_{i_1}\gamma_{i_2}}}{2\pi}\sum_{\epsilon^{(1)}_1, \epsilon^{(1)}_2}\Bigg(\frac{\braket{0|a_{i_2}|\epsilon^{(1)}_2}\braket{\bar{\epsilon}^{(1)}_2|a^\dagger_1|0}}{p_2 - \epsilon^{(1)}_2}\frac{1}{k^{\vphantom{(1)}}_1 - p^{\vphantom{(1)}}_1}\cdot \nonumber\\*
&\hspace{10em}\frac{\braket{0|a_{i_1}|\epsilon^{(1)}_1}\braket{\bar{\epsilon}^{(1)}_1|a^\dagger_1|0}}{k_1 - \epsilon^{(1)}_1}\Bigg) \nonumber\\
&+ \;\text{all permutations of $\{k_1, k_2\}$ and $\{p_1, p_2\}$}\nonumber\\
\intertext{and}
&G^{(2)}_{i_1 i_2;LL}(p_1, p_2; k_1, k_2) \nonumber\\*
= &\frac{-\i\gamma_1\sqrt{\gamma_{i_1}\gamma_{i_2}}}{2\pi}\sum_{\epsilon^{(1)}_1, \epsilon^{(2)}_2, \epsilon^{(1)}_3}\Bigg(\braket{0|a_{i_2}|\epsilon^{(1)}_3}\braket{\bar{\epsilon}^{(1)}_3|a_{i_1}|\epsilon^{(2)}_2}\cdot \nonumber \\*
&\braket{\bar{\epsilon}^{(2)}_2|a^\dagger_1|\epsilon^{(1)}_1}\braket{\bar{\epsilon}^{(1)}_1|a^\dagger_1|0}\frac{1}{k_1 - \epsilon^{(1)}_1}\frac{1}{k_1 + k_2 - \epsilon^{(2)}_2}\frac{1}{p_2 - \epsilon^{(1)}_3}\Bigg)\nonumber\\
&+ \;\text{all permutations of $\{k_1, k_2\}$ and $\{p_1, p_2\}$},
\end{align}
where it is understood that the channel $L$ ($R$) corresponds to the operator $a_1$ ($a_N$). The two-photon S-matrix elements are then written down as \cref{eq:2S} with the 4-point Green's functions in \cref{eq:2green12}. In the following discussion, we will consider the case where all the cavities are identical, i.e. $\omega_i = \omega_0, \; U_i = U \; \forall i$, with identical coupling to both waveguides, i.e. $\gamma_1 = \gamma_N = \gamma$.

\textit{Probing the dimer ($N = 2$) with two photons.} Using the calculation above, the Bose-Hubbard model with unit filling can be studied in a dimer ($N = 2$) with two photons. We study the behaviour of the resonant peaks of the S-matrix in \cref{fig:dimer} as we vary $U/J$ from the superfluid regime, $U/J \ll 1$, to the Mott insulator regime, $U/J \gg 1$, with total incoming momentum, $k_1 + k_2 = 2\omega_0 + \frac{U}{2} - \sqrt{4J^2 + \frac{U^2}{4}}$. The total incoming momentum corresponds to the state $\ket{\epsilon^{(2)}_-} \propto \ket{20} + \ket{02} - \frac{U + \sqrt{16J^2 + U^2}}{2\sqrt{2}J}\ket{11}$, which becomes the `Mott state' $\ket{11}$ as $U/J \rightarrow \infty$. As expected, when $U = 0$, the system is linear and hence the nonlinear part described by the Green's function is identically zero. When $U/J \ll 1$, the function exhibits only a single peak. As $U/J$ increases, more peaks are visible, i.e. four peaks in each quadrant, with the two off-diagonal peaks in each quadrant becoming more prominent and finally merging into one at $U/J \gg 1$. 

Several points are noteworthy. First, notice that in order for the 4-point Green's function to be identically zero at $U = 0$, either both contributions from $G^{(1)}$ ($\braket{aa^\dagger aa^\dagger}$) and $G^{(2)}$ ($\braket{aaa^\dagger a^\dagger}$) are identically zero or the contributions from $G^{(1)}$ and $G^{(2)}$ are equal and opposite. However, $G^{(1)}$ is independent of $U$ and is in general nonzero. This can be seen from \cref{fig:diagrams} where its diagram involves only the first excitation manifold that is independent of $U$. Hence, contributions from $G^{(1)}$ and $G^{(2)}$ must be equal and opposite at $U = 0$, so as to interfere destructively to give a zero 4-point Green's function.

\begin{figure}[!h]
  \centering
  \includegraphics[width=0.8\linewidth]{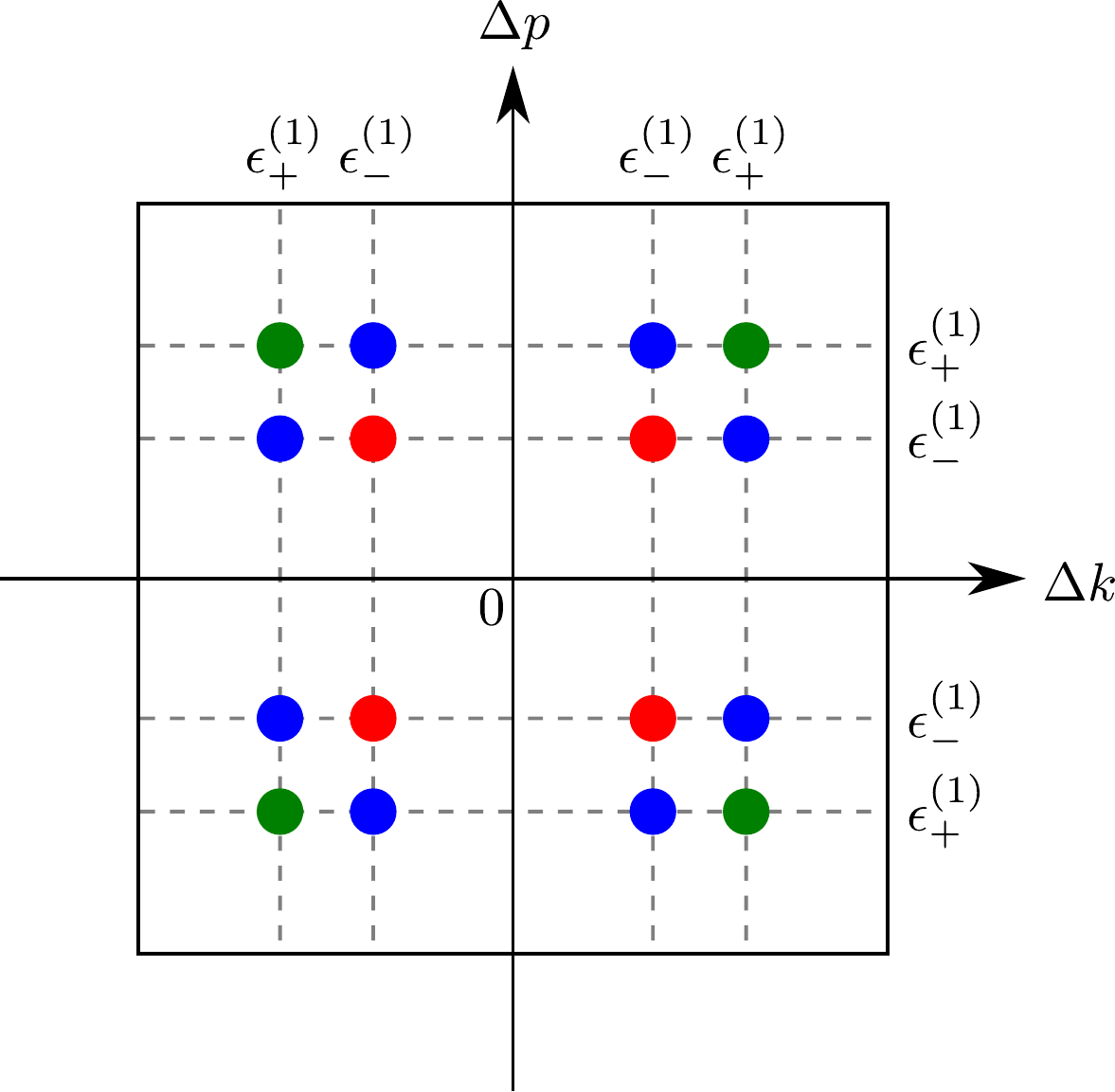}
  \caption{Positions of the peaks on the $\Delta k$-$\Delta p$ plane as determined by \cref{eq:peak}. The different colours denote different paths: red represents paths that go through $\ket{\epsilon_-^{(1)}}$ twice, blue represents paths that go through both $\ket{\epsilon_\pm^{(1)}}$ and green represents paths that go through $\ket{\epsilon_+^{(1)}}$ twice.}\label{fig:peak}
\end{figure}

Second, the positions of the peaks are determined by the following equation:
\begin{equation}\label{eq:peak}
|\Delta k (\Delta p)| = \pm 2J - \frac{U}{2} + \sqrt{4J^2 + \frac{U^2}{4}}.
\end{equation}
\Cref{fig:peak} shows a visualisation of \cref{eq:peak} together with the corresponding paths represented by different colours. 

Third, when $U/J \ll 1$, \cref{eq:peak} becomes $|\Delta k (\Delta p)| \approx 0, 4J$. However, in \cref{fig:dimer}, only one peak ($|\Delta k (\Delta p)| \approx 0$) corresponding to paths that go through $\ket{\epsilon_-^{(1)}}$ twice (red dots in \cref{fig:peak}) is visible in the transmission spectra. One can understand this by first defining symmetric and anti-symmetric operators, $a_\pm = \frac{1}{\sqrt{2}}\left(a_1 \pm a_2\right)$. Then, note that when $U/J \ll 1$ but $U \neq 0$, the contribution from $G^{(1)}$ only cancels part of the contribution from $G^{(2)}$. Hence, we can focus on the diagram represented by $G^{(2)}$, $\braket{aaa^\dagger a^\dagger}$, which is the only term that depends on $U$. The state that we are probing can be very closely approximated by the state that is created by two anti-symmetric operators:
\[\ket{\epsilon_-^{(2)}} \approx \frac{1}{\sqrt{2}}a_-^{\dagger 2}\ket{0} = \frac{1}{\sqrt{2}}a_-^\dagger\ket{\epsilon_-^{(1)}}.\]
Because of this symmetry of the system, the paths through $\ket{\epsilon_-^{(1)}}$ twice are favoured, even though the system is not pumped using the anti-symmetric operator. By choosing to pump the state, $\ket{\epsilon_-^{(2)}}$, this particular path is automatically preferred as dictated by the symmetry of the system.

Fourth, when $U/J \gg 1$, \cref{eq:peak} becomes $|\Delta k (\Delta p)| \approx \pm 2J$, making all four peak positions in each quadrant to become one, as seen in \cref{fig:dimer}. Moreover, before the four peaks in each quadrant become indistinguishable, peaks corresponding to paths that go through both $\ket{\epsilon_\pm^{(1)}}$ (blue dots in \cref{fig:peak}) are more dominant. This is because when $U/J$ is large, the symmetry described above is no longer present and different paths have similar contributions in $G^{(2)}$, however, different paths contribute differently in $G^{(1)}$ with diagram $\braket{aa^\dagger aa^\dagger}$. In $\braket{aa^\dagger aa^\dagger}$, the first excitation manifold is excited twice and hence the paths that go through both $\ket{\epsilon_\pm^{(1)}}$ are favoured since the other paths go through the same state twice which require the state to decay before it is able to be excited again.

Last, in the intermediate regime where $U/J \sim 1$, the behaviour in \cref{fig:dimer} can be understood as an interplay and continuation between the two extreme regimes. Further note that the height of the peaks representing the strength of the nonlinearity increases with $U$ as expected.

\begin{figure*}[!htb]
  \centering
  \includegraphics[width=\textwidth]{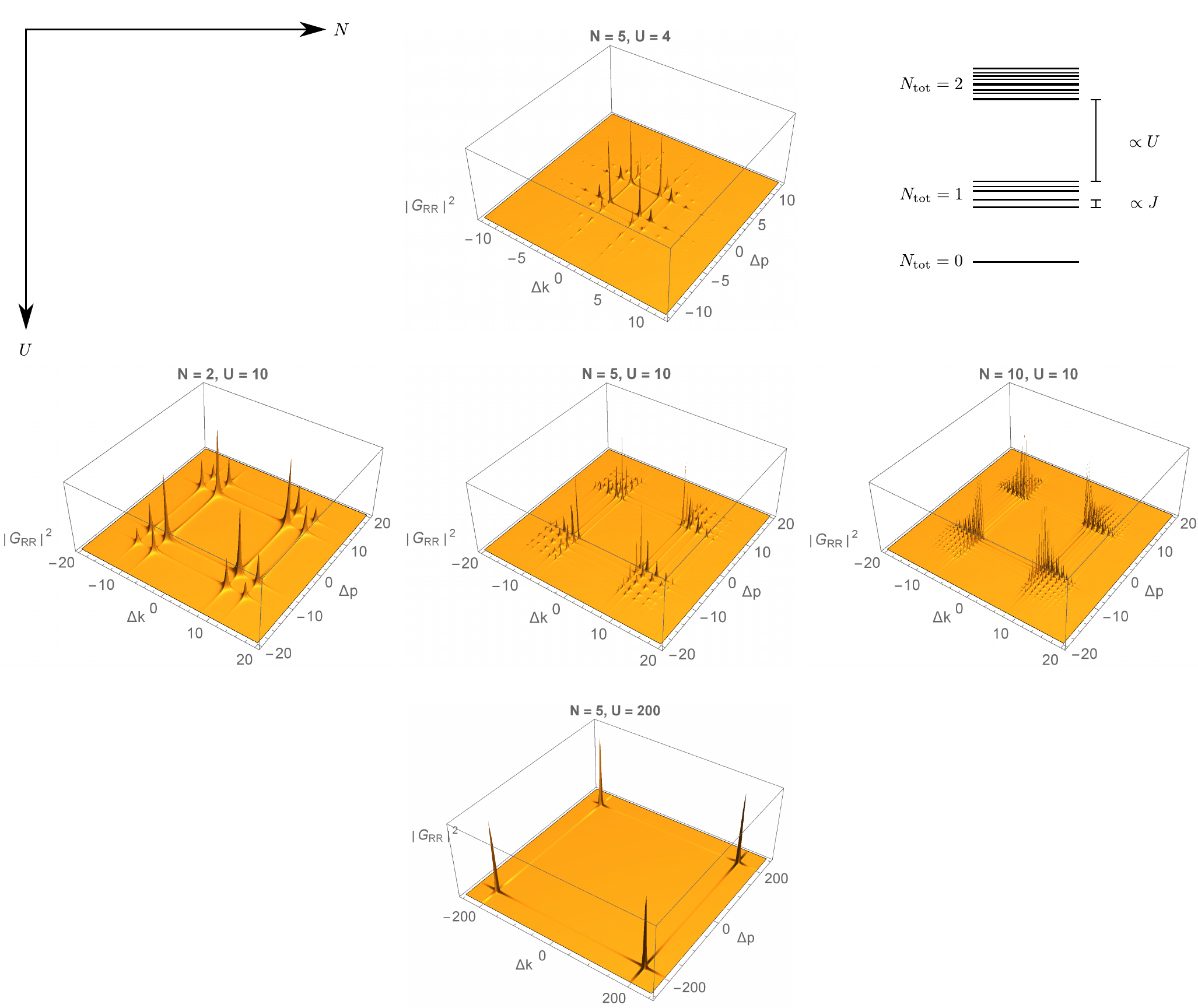}
  \caption{\textbf{Bose-Hubbard model:} Plots of $|G_{RR;LL}|^2$ (incoming photon channel labels $LL$ are dropped) of the Bose-Hubbard model with the open boundary condition. $N = 2, 5, 10$ and $U = 4, 10, 200$ are shown. $\Delta k \equiv k_1 - k_2$ and $\Delta p\equiv p_1 - p_2$. $\omega_0 = 100$, $\gamma = 0.25$ and the total incoming momentum, $k_1 + k_2 = \text{highest doubly excited eigenvalue}$, all in units of $J$. The schematic diagram at the top right corner depicts the scaling of the level spacings.}\label{fig:bhmgrr}
\end{figure*}

\begin{figure*}[!ht]
  \centering
  \includegraphics[width=0.32\textwidth]{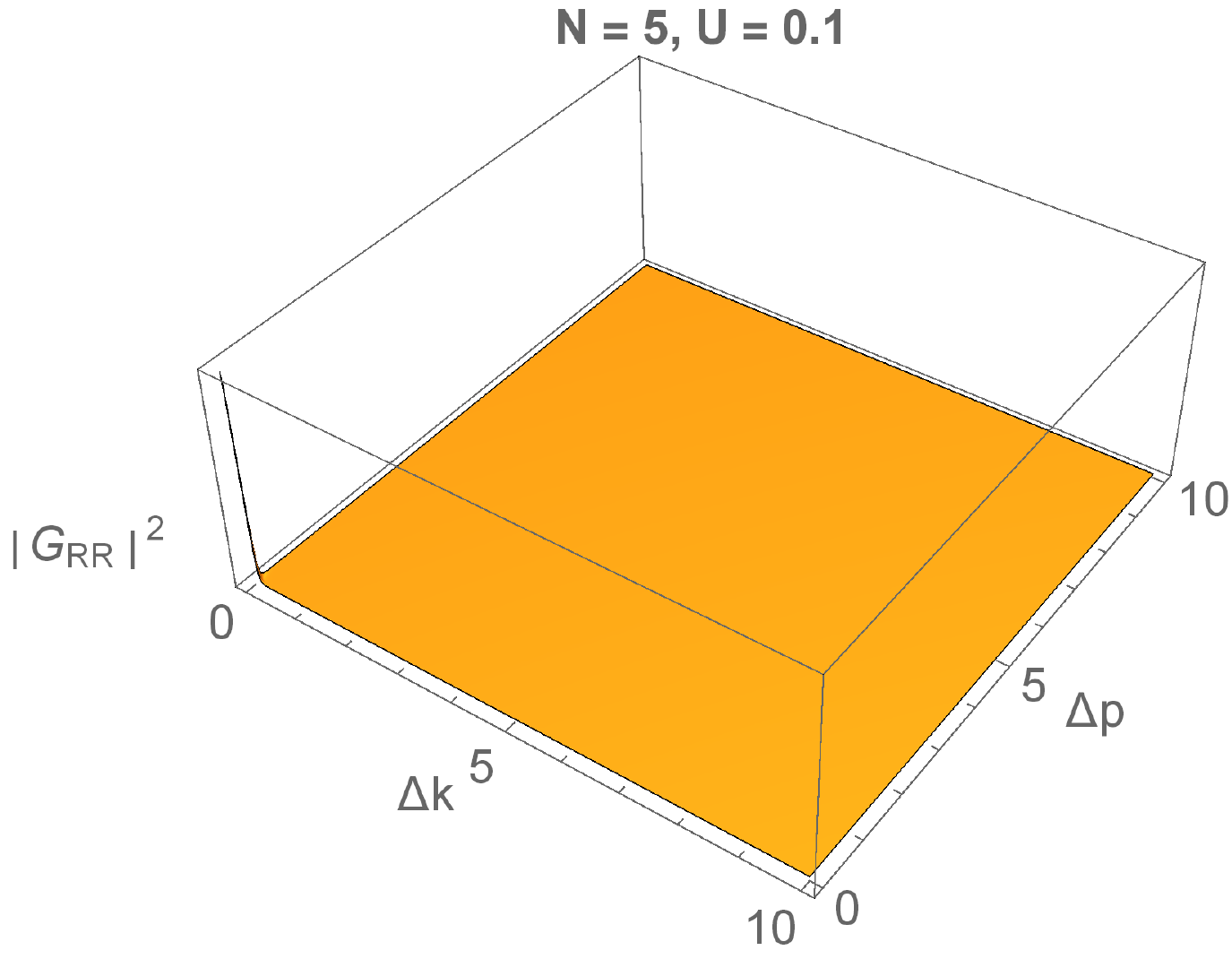}\qquad
  \includegraphics[width=0.32\textwidth]{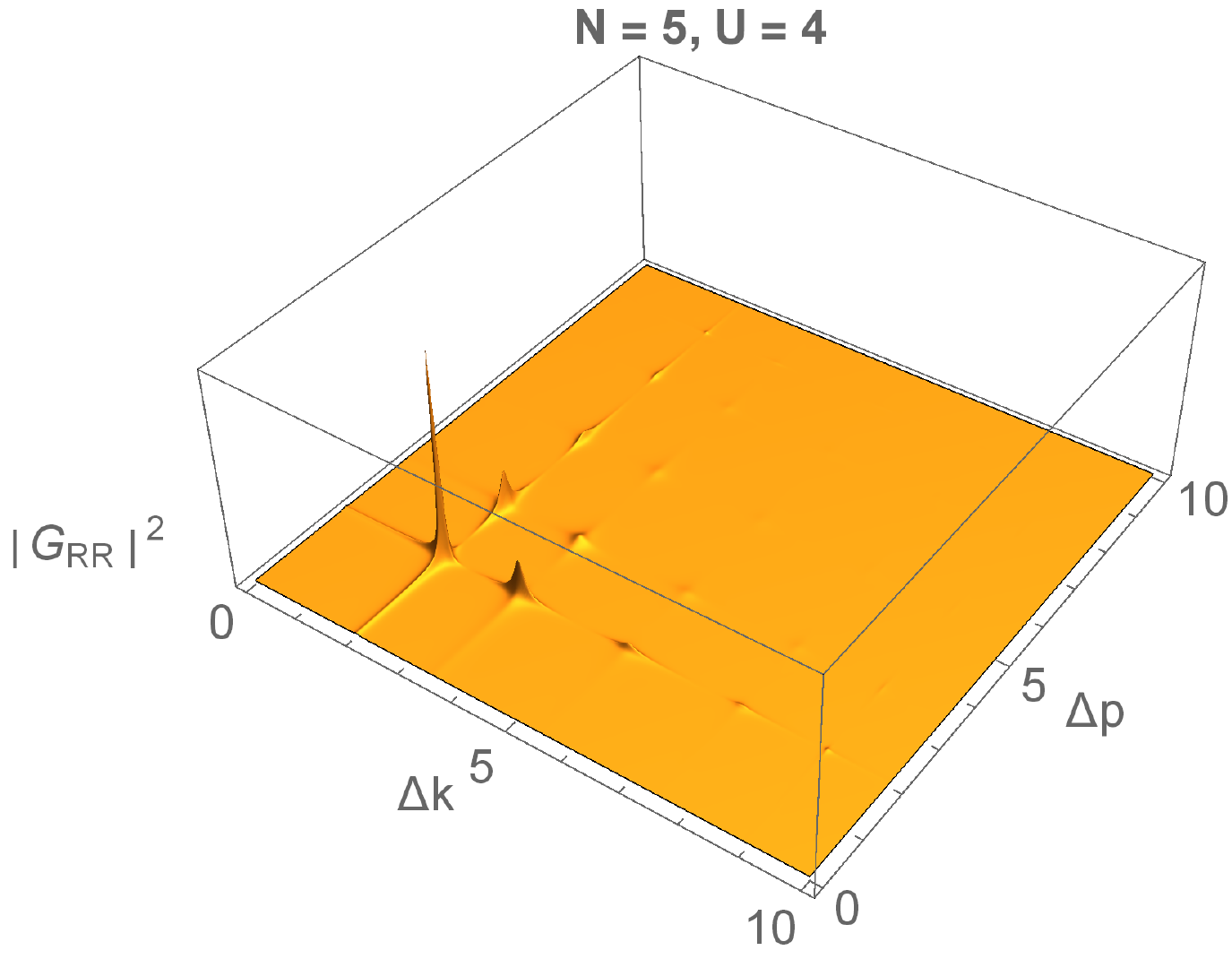}\\
  \includegraphics[width=0.32\textwidth]{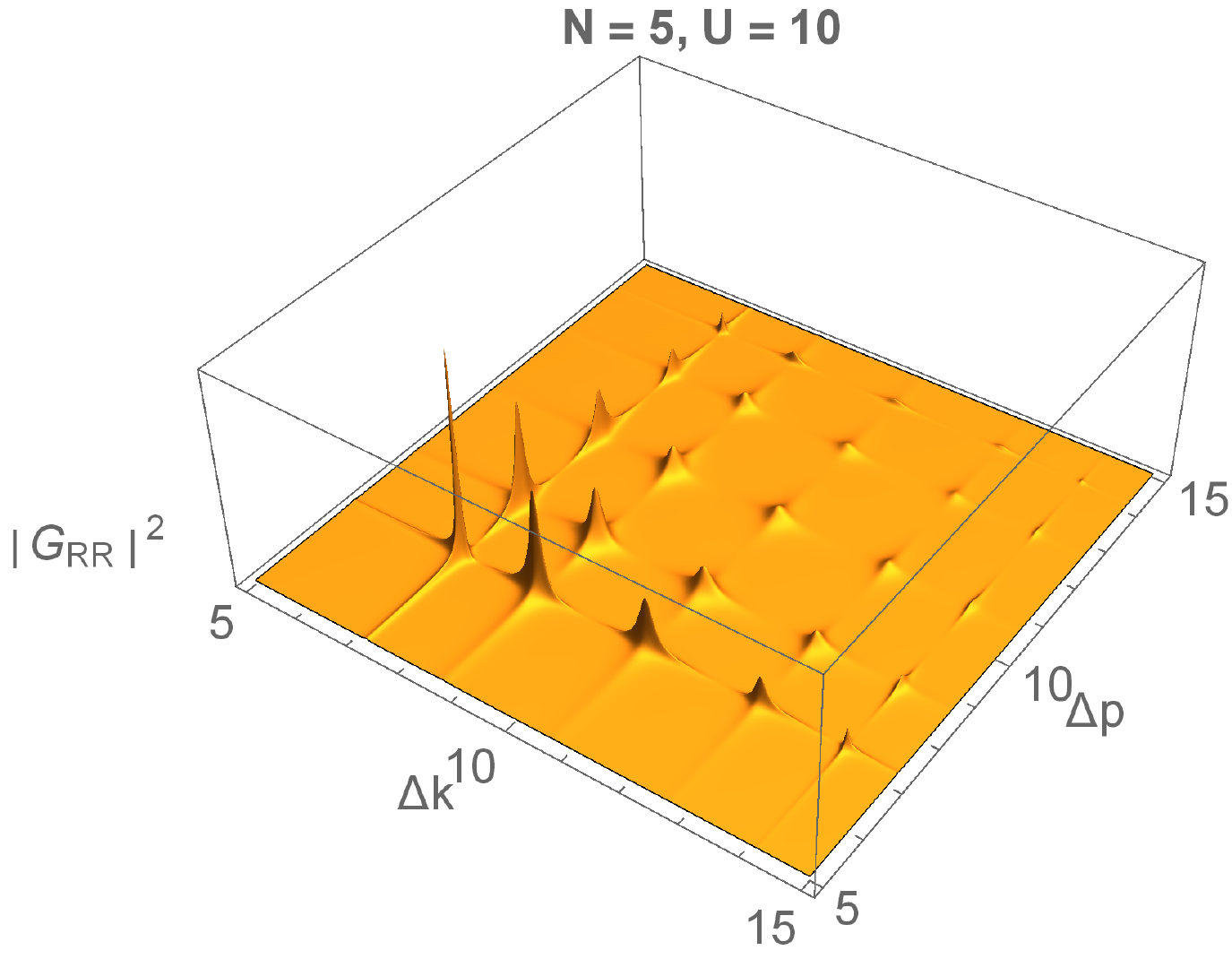}\qquad
  \includegraphics[width=0.32\textwidth]{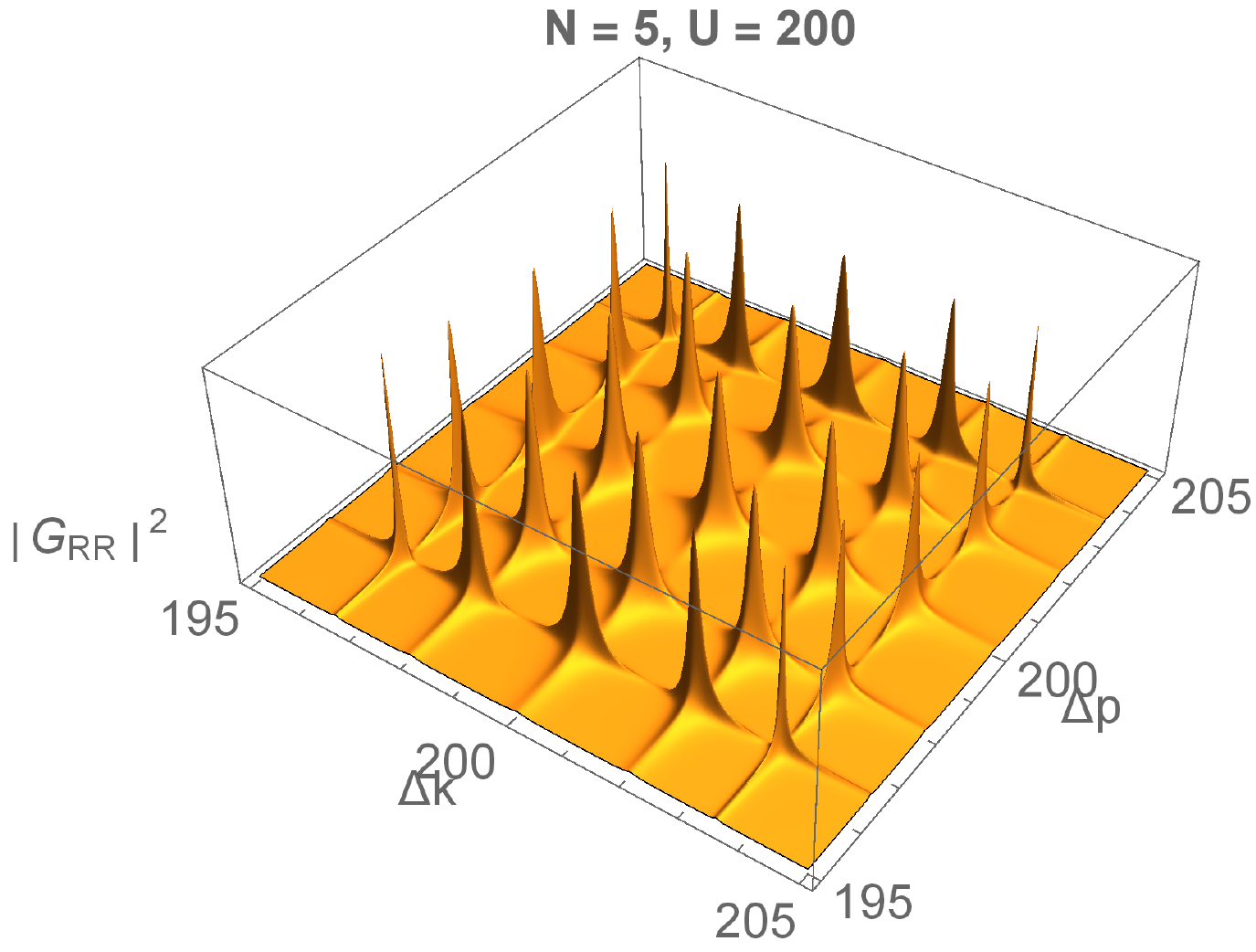}
  \caption{\textbf{Bose-Hubbard model with $N = 5$:} Plots of $|G_{RR;LL}|^2$ (incoming photon channel labels $LL$ are dropped) of the Bose-Hubbard model with the open boundary condition. $N = 5$ and $U = 0.1, 4, 10$ and $200$ respectively. $\Delta k \equiv k_1 - k_2$ and $\Delta p\equiv p_1 - p_2$. $\omega_0 = 100$, $\gamma = 0.25$ and the total incoming momentum, $k_1 + k_2 = \text{highest doubly excited eigenvalue}$, all in units of $J$.}\label{fig:bhmu}
\end{figure*}

\textit{Many sites regime.} Finally, $|G_{RR;LL}|^2$ is plotted in \cref{fig:bhmgrr,fig:bhmu} with different numbers of cavities, $N$ and different nonlinearities, $U$ at a total energy of two incoming photons, $k_1 + k_2$ equal to the highest doubly excited state. In \cref{fig:bhmgrr}, off-diagonal peaks ($|\Delta k| \neq |\Delta p|$) that can only result from nonlinear scattering processes are clearly visible. Also, the positions of peaks in \cref{fig:bhmgrr} correspond to the eigenenergies in the single excitation manifold. Hence, as $N$ increases, the number of peaks increases. \Cref{fig:bhmu} focuses on the positive quadrant with $N =5$ and varying $U$, going from the weakly interacting to strongly interacting regimes. We note three features: with increasing nonlinearity, i) the peaks are located at progressively larger values (refer to \cref{fig:bhmgrr} in which the clusters of peaks move further apart as $U$ increases); ii) the heights of the peaks become more and more uniform; iii) the relative spacings between the peaks are maintained and when $U/J \ll 1$ only one peak is visible. These features reflect the following facts: i) the highest doubly excited state has a larger energy as the nonlinearity increases; ii) the overlaps between the highest doubly excited state with $a_1\ket{\epsilon^{(1)}}$ and $a_N\ket{\epsilon^{(1)}}$ become more uniform as the nonlinearity increases; iii) the spacings between the peaks reflect the relative gaps between the singly excited states which are independent of $U$ and the single peak at $U/J \ll 1$ can be explained by the same symmetry argument as in the case of two cavities. 

\section{Conclusion}\label{sec:conclusion}
In summary, we have presented a diagrammatic approach to construct the Green's functions (and in turn the S-matrix) required to study multi-photon transmission in one dimensional systems comprised of correlated quantum emitters. Our method can be used for any bosonic many-body systems probed by corresponding leads. The method is especially useful in working out inelastic parts of the scattering matrix. We have demonstrated the usefulness of our technique by going through the calculations of the S-matrix for a few paradigmatic examples including cases of single and many correlated emitters. Through them, we showed that the method simplifies calculations considerably, especially when we are dealing with interacting many-body problems, even at the two-photon level. We also discussed the applicability of our method when one deals with multi-photons scattering from non-interacting emitters, by providing an intuitive way to visualise the scattering processes that map one-to-one to Green's functions. Having the ability to write down expressions of the Green's functions directly and to visualise them will be useful in guessing a system's response qualitatively without computing the exact form. 

\section{Acknowledgements}
This research is supported by Singapore Ministry of Education Academic Research Fund Tier 3 (Grant No. MOE2012-T3-1-009),  National Research Foundation (NRF) Singapore and the Ministry of Education, Singapore under the Research Centres of Excellence programme.

\appendix
\section{Cluster decomposition of the S-matrix}\label{sec:clusterS}
\begin{figure*}
\centering
\includegraphics[width=0.8\textwidth]{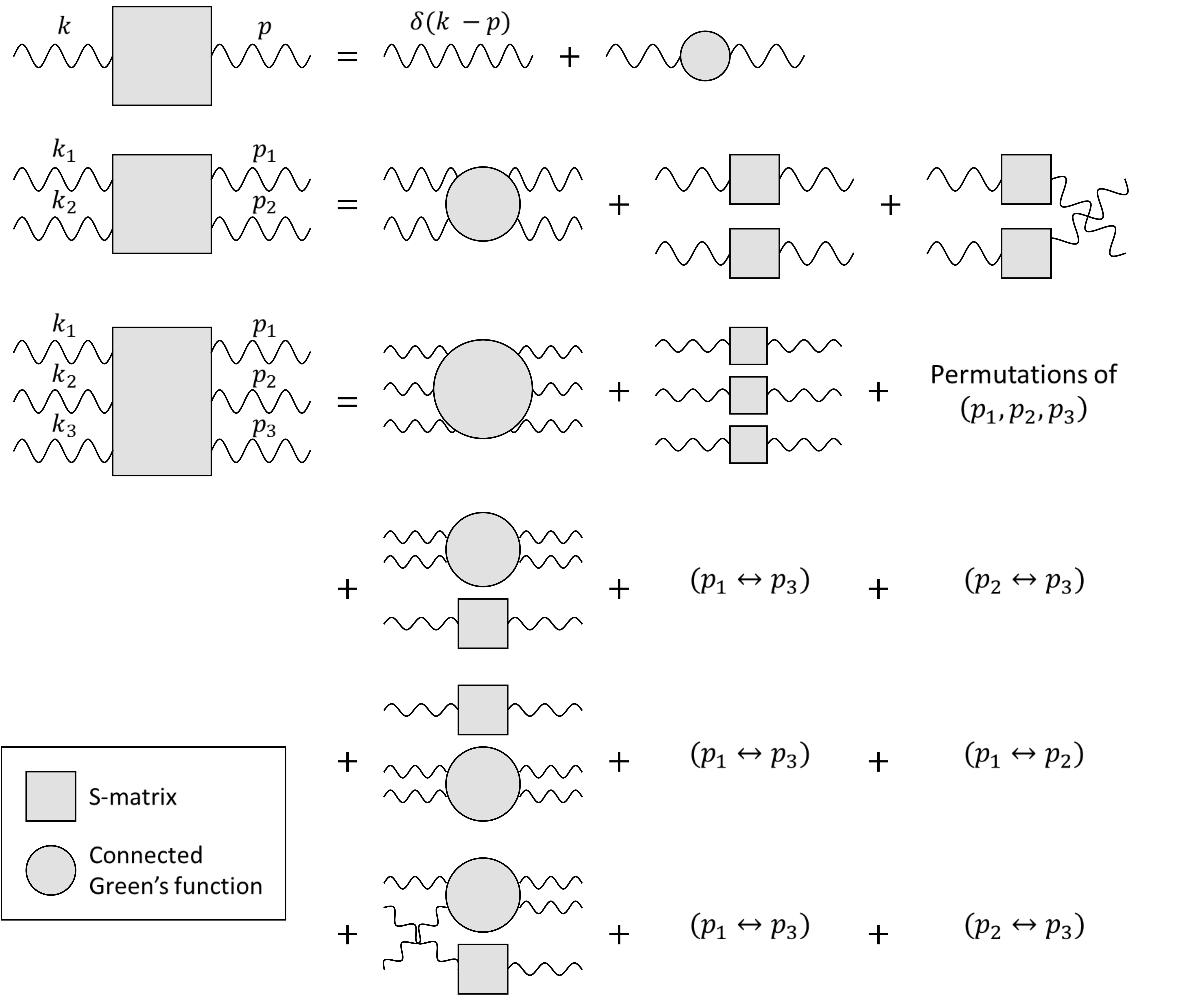}\\
\caption{Cluster decomposition structure of the one-, two-, and three-photon  S-matrix. The $n$-photon S-matrix is made up of $2m$-point connected Green's functions, for $m \leqslant n$. The delta function in the one-photon S-matrix is present only if the input from the corresponding waveguide is nonzero.}\label{fig:clusterS}
\end{figure*}

Figure \ref{fig:clusterS} illustrates the cluster decomposition structure of the S-matrix for up to three photons. We can see that an elastic scattering, in which the photon momenta are simply rearranged at the output, is fully described by the single-photon S-matrix and a larger-photon-number elastic scattering amplitude is just the product of it. This, however, only describes the linear part of the system's response. If the system is nonlinear, there will be inelastic scatterings as well, which gives the S-matrix a richer structure. This is encoded in the Green's functions. Therefore, in order to see the effects of nonlinearity, we need to calculate the Green's functions which form the important part of the S-matrix.

\section{Proof}\label{sec:proof}
Here, we will prove the steps outlined in \cref{sec:diagram}. The time ordering in \cref{eq:gfun} gives rise to different diagrams. To see this, consider a particular time ordering, $\braket{a\dots aa^\dagger\dots a^\dagger}$, up to interchanges of the $a$'s or $a^\dagger$'s at different times that do not change the form. Then, we insert identities, $\mathds{1} = \sum_\epsilon\ket{\epsilon}\bra{\bar{\epsilon}}$ in between two operators to evaluate the expression.
\begin{align}\label{eq:1eva}
&\braket{0|a(t'_1)\dots a(t'_m)a^\dagger(t_1)\dots a^\dagger(t_m)|0} \nonumber\\*
&\cdot\theta(t'_1 - t'_2)\dots\theta(t_{m - 1} - t_m)(-1)^m \nonumber \\*
= &\sum_{\epsilon_1 \dots \epsilon_{2m - 1}} \braket{0|a(t'_1)|\epsilon_{2m - 1}}\braket{\bar{\epsilon}_{2m - 1}|\dots|\epsilon_1}\braket{\bar{\epsilon}_1|a^\dagger(t_m)|0} \nonumber\\*
&\hspace{4em}\cdot\theta(t'_1 - t'_2)\dots\theta(t_{m - 1} - t_m)(-1)^m \nonumber\\
= &\sum_{\epsilon_1 \dots \epsilon_{2m - 1}} \braket{0|a|\epsilon_{2m - 1}}\braket{\bar{\epsilon}_{2m - 1}|\dots|\epsilon_1}\braket{\bar{\epsilon}_1|a^\dagger|0} \nonumber\\*
&\hspace{4em}\cdot\e^{-\i\epsilon_{2m - 1}(t'_1 - t'_2)}\dots\e^{-\i\epsilon_1(t_{m - 1} - t_m)} \nonumber\\*
&\hspace{4em}\cdot\theta(t'_1 - t'_2)\dots\theta(t_{m - 1} - t_m)(-1)^m
\end{align}
where we have taken just one particular variant of the form $\braket{a\dots aa^\dagger\dots a^\dagger}$. To find the Green's function in the momentum space, we perform a Fourier transformation of \cref{eq:1eva}. Since only the exponential and Heaviside step functions are time-dependent, all we need to do is
\begin{multline}\label{eq:proof}
\mathscr{F}\Big(\e^{-\i\epsilon_{2m - 1}(t'_1 - t'_2)}\dots\e^{-\i\epsilon_1(t_{m - 1} - t_m)}\\* \theta(t'_1 - t'_2)\dots\theta(t_{m - 1} - t_m)(-1)^m\Big).
\end{multline}
To calculate this, we perform a change of variables,
\begin{align*}
\tau_i &= t_{m - i} - t_{m - i + 1} &\text{for}\;i &= 1, \dots, m - 1 \nonumber\\
\tau_{m} &= t'_m - t_1 && \nonumber\\
\tau_{m + i} &= t'_{m - i} - t'_{m - i + 1} &\text{for}\;i &= 1, \dots, m - 1 \nonumber\\
\tau_{2m} &= t_m &&
\end{align*}
and define
\begin{align*}
\vec{t} &= \begin{pmatrix}
	t_m \\
	\vdots \\
	t_1 \\
	t'_m \\
	\vdots \\
	t'_1
\end{pmatrix}, 
&\bm{\alpha} &= \begin{pmatrix*}[r]
	k_m \\
	\vdots \\
	k_1 \\
	- p_m \\
	\vdots \\
	- p_1
\end{pmatrix*}, \nonumber\\
\bm{\tau} &= \begin{pmatrix}
	\tau_1 \\
	\vdots \\
	\tau_{2m}
\end{pmatrix},
&M &= \begin{pmatrix*}[r]
	0 & 0 & 0 & \hdots & 0 & 1 \\
	1 & 0 & 0 & \hdots & 0 & 1 \\
	1 & 1 & 0 & \hdots & 0 & 1 \\
	\vdots & \vdots & \vdots & \ddots & \vdots & \vdots \\
	1 & 1 & 1 & \hdots & 1 & 1
\end{pmatrix*}\\
\end{align*}
where $\vec{t} = M\bm{\tau}$ and $\mathscr{F} = \int\frac{\dd^{2m}\vec{t}}{(2\pi)^m}\;\e^{-\i\bm{\alpha}.\vec{t}}$. \Cref{eq:proof} now becomes
\begin{align*}
&\left(\frac{-1}{2\pi}\right)^m\left(\prod_{i = 1}^{2m - 1}\int_0^\infty\dd\tau_i\int_{-\infty}^\infty\dd\tau_{2m}\right)\e^{-\i\bm{\alpha}.\vec{t}}\prod_{j = 1}^{2m -1}\e^{-\i\epsilon_j\tau_j} \\*
= &\left(\frac{-1}{2\pi}\right)^m\left(\prod_{i = 1}^{2m - 1}\int_0^\infty\dd\tau_i \e^{-\i(\sum_{j = i + 1}^{2m}\alpha_j + \epsilon_i)\tau_i}\right)\cdot \\*
&\hspace{3em}\left(\int_{-\infty}^\infty\dd\tau_{2m}\e^{-\i\sum_{j = 1}^{2m}\alpha_j\tau_{2m}}\right)\\
= &\frac{(-1)^m}{(2\pi)^{m - 1}}\;\delta\left(\sum_{j = 1}^{2m}\alpha_j\right)\left(\prod_{i = 1}^{2m - 1}\frac{-\i}{\sum_{j = i + 1}^{2m}\alpha_j + \epsilon_i}\right) \\
= &\frac{-\i}{(2\pi)^{m - 1}}\;\delta\left(\sum_{i = 1}^m k_i - \sum_{i = 1}^m p_i\right)\left(\prod_{i = 1}^{2m - 1}\frac{1}{\sum_{j = 1}^i\alpha_j - \epsilon_i}\right)
\end{align*}
The other variants in $\braket{a\dots aa^\dagger\dots a^\dagger}$ are accounted for by permutations of the $k$'s and $p$'s of the previous line. This together with the sum over weights in \cref{eq:1eva} gives us all the factors detailed in \cref{sec:diagram}. If instead we have a different time ordering, the weights that we sum over in \cref{eq:1eva} will be different but will take a straight forward form. Moreover, the Heaviside step functions in \cref{eq:proof} will be such that they  show the time ordering together with the exponential functions. For example, if we consider an ordering like $\braket{aa^\dagger aa^\dagger \dots aa^\dagger}$, the weights to sum over will look like
\[\sum_{\epsilon_1 \dots \epsilon_{2m - 1}} \braket{0|a|\epsilon_{2m - 1}}\braket{\bar{\epsilon}_{2m - 1}|a^\dagger\dots a|\epsilon_1}\braket{\bar{\epsilon}_1|a^\dagger|0}\]
and \cref{eq:proof} will look like
\begin{multline}
\mathscr{F}\Big(\e^{-\i\epsilon_{2m - 1}(t'_1 - t_1)}\e^{-\i\epsilon_{2m - 2}(t'_2 - t_2)}\dots\e^{-\i\epsilon_1(t'_m - t_m)}\\ \theta(t'_1 - t_1)\theta(t'_2 - t_2)\dots\theta(t'_m - t_m)(-1)^m\Big)
\end{multline}
up to permutations of $a$'s and $a^\dagger$'s at different times. The proof can then be proceeded as before by defining $\bm{\tau}$ in a similar fashion, and $\vec{t}$ as a list from earliest time to latest time and $\bm{\alpha}$ accordingly. It is now obvious to see that different time orderings give the same general expression but can be represented with different diagrams. For non-particle number conserving Hamiltonians, the same expression holds but the sum will now have to be taken over all eigenenergies instead of a particular manifold and the interpretation of the diagrams will be different. 

\section{4-point Green's function for two collocated atoms}\label{sec:exact2coll}
The exact expression of the 4-point Green's function for two collocated atoms coupled to a waveguide (\cref{sec:2coll}) is as below: 
\begin{align*}
&G(p_1, p_2; k_1, k_2) \nonumber\\*
= &\frac{\i}{32\pi\gamma_c^2}\frac{E_i - 2\omega_c}{E_i - 2\omega_c + \i\gamma_c}g_{k_1}g_{k_2}g_{p_1}g_{p_2}\Bigg(4(E_i - 2\omega_c + 2\i\gamma_c) \nonumber\\
&\hspace{5em}+ \omega_d^2 f^{(2)}(\vec{p};\vec{k}) + \omega_d^4 f^{(4)}(\vec{p};\vec{k})\Bigg)\delta(E_i - E_o)
\end{align*}
where 
\begin{align*}
f^{(2)}(\vec{p};\vec{k}) &= \frac{(E_i - 2\omega_c)\Big(\splitfrac{(k_1 - k_2)^2 + (p_1 - p_2)^2}{+ 4\gamma_c^2 + 2(E_i - 2\omega_c + 2\i\gamma_c)^2}\Big)}{(k_1 - \omega_c)(k_2 - \omega_c)(p_1 - \omega_c)(p_2 - \omega_c)} \\
f^{(4)}(\vec{p};\vec{k}) &= -\frac{4(3(E_i - 2\omega_c) + 2\i\gamma_c)}{(k_1 - \omega_c)(k_2 - \omega_c)(p_1 - \omega_c)(p_2 - \omega_c)}. 
\end{align*}

\bibliography{BibEIT}

\end{document}